\documentclass[aps,prb,showpacs,amsmath,amssymb,reprint,10pt,superscriptaddress,longbibliography]{revtex4-2}
\usepackage{hyperref}
\usepackage[utf8]{inputenc}
\usepackage{url}
\usepackage{pifont}
\usepackage[english]{babel}
\usepackage[utf8]{inputenc}
\usepackage{graphicx,epsfig,xcolor}
\usepackage{latexsym}
\usepackage{amsfonts}
\usepackage{amsmath}
\usepackage{placeins}
\onecolumngrid
\usepackage{polski}
\usepackage{braket}
\usepackage[bottom]{footmisc}
\usepackage[normalem]{ulem}
\usepackage{graphicx} 
\usepackage{subfigure}

\def\ket|#1>{| #1 \rangle}
\def\bra<#1|{\langle #1 |}
\def\<{\langle}
\def\>{\rangle}
\def\{{\lbrace}  
\def\}{\rbrace}
\def\beq{\begin{equation}}
\def\eeq{\end{equation}}

\begin{document}

\title{Quantum Complexity in Rule-Based Constrained Many-Body Models:\\ Scars, Fragmentation, and Chaos}
\author{Arkaprava Sil}
\email{24dr0041@iitism.ac.in}
\author{Sudipto Singha Roy}
\email{sudipto@iitism.ac.in}
\affiliation{Department of Physics, Indian Institute of Technology (Indian School of Mines) Dhanbad, IN-826004, Dhanbad, India}
\date{\today}

\begin{abstract}
Kinetic constraints in quantum many-body systems strongly restrict the accessible Hilbert space, giving rise to highly nontrivial dynamical behavior. In recent years, such systems have attracted growing interest as they provide insight into mechanisms of thermalization and into regimes where thermalization fails. In this work, we study a family of \textit{rule-based} kinetically constrained models, including the celebrated Quantum Game of Life, from the perspective of quantum complexity, with a focus on entanglement, nonstabilizerness, and quantum chaos.
Using spectral diagnostics such as level statistics and spectral form factors, we show that these models exhibit robust chaotic behavior while simultaneously supporting both strong and weak Hilbert-space fragmentation and quantum many-body scar states. Our results reveal that the nature of Hilbert-space fragmentation can be tuned qualitatively from strong to weak fragmentation and ultimately to its absence, within a single family of models through simple variations of the underlying kinetic rules. To further elucidate the structure of these fragmented subspaces, we characterize them through their ability to generate quantum resources. In particular, we show that resource-generation capacity does not necessarily correlate with the dimensionality of a fragmented sector, and that entanglement structure and the ability to generate nonstabilizerness provide effective diagnostics for distinguishing dynamically disconnected sectors, including those supporting nonthermal scarred dynamics. Our work thus places kinetically constrained models within a broader and more general framework, not limited to Rydberg blockade based constraints only, and demonstrates that simple variations in underlying rules can lead to qualitatively distinct static and dynamical regimes including chaotic, fragmented, and scarred phases that remain largely unexplored in previous works that can be instrumental both for fundamental understanding and quantum technological applications of these models.

\end{abstract}
\maketitle

\section{Introduction}
Over the past decades, understanding thermalization in closed quantum systems has been a key question in quantum many-body physics. It is well known that integrable systems do not thermalize in the conventional sense, and instead relax to generalized Gibbs ensembles. This naturally leads to the question: {\it do all non-integrable systems thermalize}? One possible explanation for thermalization comes from the eigenstate thermalization hypothesis (ETH)~\cite{ETH_ansatz1,ETH_ansatz2,ETH_ansatz3,ETH_ansatz4,ETH_ansatz5,ETH_ansatz6,ETH_ansatz7}, but there are notable exceptions~\cite{Nandkishore_2017,ALET2018498, Abanin_2019,Sierant_2025}. Among the disorder-free spatially homogeneous quantum many-body models, kinetically constrained models (KCMs) often contain quantum states with unusual behavior that do not thermalize, commonly known as the  quantum many-body scar (QMBS) states~\cite{Yao_2025,Garcia_2026,Deger_2024,Turner_2018_NatPhys,Turner_2018_PRB, Serbyn_2021,Iadecola_2020,Wang_2024,Hu_2025,Zhang_2023}. These scars offer a clear example of non-thermal behavior and show lower entanglement entropy. Recent experiments on a 51-atom quantum simulator provide direct evidence of quantum many-body scars~\cite{Bernien_2017}. See also Refs.~\cite{Valado2016,Zhang2023HilbertScar,su2023} for other recent experimental implementations of QMBS.
Moreover, recent studies have found another route to non-ergodicity through Hilbert space fragmentation (HSF)~\cite{Sala2020,Mukherjee2021,Moudgalya_2022,Kohlert2023,Anwesha2023,Francica2023,Somsubhra_2023, Somsubhra_2024, Aditya_2024,Maitri_2025}.  In these systems, the Hilbert space splits into dynamically disconnected sectors, which prevents full thermalization.  Experimental examples of fragmentation have already been shown in platforms such as cold atoms and Rydberg arrays~\cite{Kohlert2023, Scherg2021, Adler2024,yang_2025,Wang_2025}.

In recent years, much of the interest in studying kinetically constrained models has been driven by their experimental realization in Rydberg atom arrays operating in the blockade regime. This experimental accessibility has opened avenues to explore a broader range of physical phenomena. For instance, while the PXP model is known to host quantum many-body scar states, it does not exhibit Hilbert-space fragmentation; such fragmentation has been shown to emerge in Floquet generalizations of the PXP model~\cite{Mukherjee2021}, exhibiting coexistence of both scar and fragmentation. More recently, related forms of constrained dynamics have also been explored in the antiblockade regime of Rydberg systems, where facilitation mechanisms lead to distinct dynamical rules and constraint structures~\cite{Zhao_2025}. 

In parallel, recent years have also witnessed growing interest in kinetically constrained quantum dynamics formulated  in terms of  \emph{rule-based} evolution, commonly known as ``quantum cellular automata"~\cite{Farrelly_2020}, where local updates depend on the population of neighboring sites rather than on a fixed symmetric projector structure. A prominent example is the Quantum Game of Life (QGL)~\cite{bleh2012qgol,Arrighi2012,ney2022life}, introduced as a quantum analogue of John Conway’s classical Game of Life~\cite{gardner1970life}, in which the dynamics at a given site is governed by the total occupation of its surrounding sites. 

Motivated by the rule-based perspective exemplified by the Quantum Game of Life, and using symmetric projector-based PXP-type models as reference points, we focus in this work on a  family of one-dimensional kinetically constrained quantum many-body models in which the dynamics at site $i$ depends on the total population of its neighboring sites, including both nearest neighbors $(i-1,i+1)$ and next-nearest neighbors $(i-2,i+2)$. When expressed in operator form, such population-dependent dynamical rules naturally give rise to sum of  \emph{asymmetric projectors}, in contrast to the symmetric constraints appearing in standard PXP or PPXPP~\cite{Cheng_2023,Kerschbaumer_2025}. In other words, this provides a route to explore kinetic constraints beyond strict blockade conditions, leading to qualitatively different dynamical structures.

We begin by examining the role of kinetic constraints in the emergence of the chaotic behavior of the considered KCMs. Despite earlier investigations~\cite{Civera2021}, it remains unclear whether the Quantum Game of Life and related kinetically constrained models exhibit genuine quantum chaotic behavior.
Using standard chaos diagnostics such as level-spacing statistics and the spectral form factor (SFF), we demonstrate that, along with QGL, all variants of kinetically constrained models considered in this work display clear signatures of quantum chaos when analyzed systematically. A key observation is that symmetry resolution plays a crucial role in revealing these features. However, for a subset of the models, resolving symmetries alone is insufficient: the spectrum further exhibits Hilbert space fragmentation~\cite{Haowei_2026}, requiring an analysis at the level of dynamically disconnected sectors.  The chaotic behavior of these models becomes evident only when these sectors are probed separately.
Our analysis further reveals  that Hilbert space fragmentation (HSF) depends strongly on the type of constraint used, revealing weak, moderate, or very strong fragmentation within a {\it single} family of KCMs  we considered here. Interestingly, in some cases, even the symmetry-resolved fragmented sectors exhibit quantum many-body scars with varying stability~\cite{Nandy_2024}.

We next characterize these chaotic subspaces from the perspective of quantum resource generation, focusing on entanglement and non-stabilizerness. In this context, entanglement and non-stabilizerness provide us with some insight regarding  state-preparation complexity (see also~\cite{Salvatore_2023,Soumik_2023, Haug_2025, Turkeshi_2025}). In literature, there have been prior studies on such aspects. For example, a very recent work by Santra \textit{et al.}~\cite{santra_2025} discusses quantum complexity in disordered models. Among the KCMs, Refs.~\cite{Yuan_2022, smith_2025,Ivanov_2025} explore the quantum complexity properties of the PXP model and its variants~\cite{Sierant_2021,Kerschbaumer_2025}. In our case, this characterization is particularly important in the presence of Hilbert space fragmentation: unlike symmetry-resolved sectors, fragmented subspaces cannot be identified through conventional conserved quantities. Instead, we probe them using complexity-based measures. Specifically, we evaluate the entanglement entropy~\cite{Horodecki_2009} and the stabilizer R{\'e}nyi entropy~\cite{Leone_2022} for the largest connected component as well as several higher-dimensional fragmented subspaces, and compare their capacity for resource generation across dynamically disconnected sectors. We find that higher-dimensional fragmented subspaces do not necessarily correspond to the most resource-generating sectors, indicating that size alone does not determine resource content.  In this way, just as conserved quantities distinguish symmetry sectors, complexity measures allow us to differentiate fragmented subspaces through their static and dynamical resources.

Taken together, our results demonstrate that even structurally similar kinetically constrained models can exhibit a wide range of nontrivial physical behavior. To the best of our knowledge, this diversity has not been reported in prior studies. By showing that quantum many-body scars, a variety of Hilbert-space fragmentations, and slow dynamics persist in systems governed by population-dependent rules, rather than symmetric projector-based constraints, we establish that these phenomena are not tied  to a specific Hamiltonian form (such as those arising from Rydberg blockade). Instead, they emerge more generally from the underlying constraint structure, thereby broadening the theoretical framework of kinetically constrained quantum dynamics. Beyond their fundamental significance, our results may also be relevant for emerging quantum technologies, particularly in light of recent proposals that exploit kinetically constrained models for applications such as quantum sensing~\cite{Dooley_2021,yoshinaga2022}.  We reiterate that, although several works have addressed quantum cellular automata and other rule-based models, a systematic and unified characterization of different notions of quantum complexity in such systems has so far been lacking that we addressed in our work.  We stress here that in this work, we use the term \emph{quantum complexity} specifically in the context of quantum state preparation and the simulation of its dynamics. In the broader literature, however, quantum complexity often refers to notions from quantum complexity theory and complexity classes~\cite{watrous2008quantum}, which lie beyond the scope of the present article.

This work is organized as follows. In Section \ref{Model Hamiltonians}, we introduce the family of kinetically constrained models that we examine in this study. Following this, in Section \ref{Sec:Hamiltonian-Induced Hilbert Space Structure}, we discuss the effect of these kinetic constraints on the dimension of the effective Hilbert space, including how they lead to Hilbert space fragmentation and their associated symmetries. In Section \ref{Sec:quantum complexity}, we provide a detailed analysis of the quantum complexity aspects of these models, exploring entanglement, stabilizer R{\'e}nyi entropy and  quantum chaos. Sec.~\ref{sec:quantum scars} further discusses the appearance of quantum many-body scar states in the dynamically disconnected fragmented sector and their robustness. The quantum complexity properties of the Hamiltonian $H^{\mathrm{Tot}}$ are discussed separately in Sec.~\ref{sec:Htot}, highlighting key aspects that differ from those observed in other models.
Finally, in Section \ref{Sec discussion}, we present a discussion on the future outlook of our work.

\section{Model Hamiltonians}
\label{Model Hamiltonians}
Before we present the details of our analysis, in this section, we define the set of rule-based quantum many-body Hamiltonians that we have considered in our work. One such primary candidate is the Hamiltonian $H^{\mathrm{Tot}}$, acting on a 1D spin-1/2 chain, which we define below. 
\begin{align}
    H^{\mathrm{Tot}} &= \sum_{i=1}^L \sigma^x_i (\mathcal{N}_i^{(0)}+\mathcal{N}_i^{(1)}+\mathcal{N}_i^{(2)} + \mathcal{N}_i^{(3)}), \nonumber\\
    &=H^{\mathcal{N}(0)}+H^{\mathcal{N}(1)}+H^{\mathcal{N}(2)}+H^{\mathcal{N}(3)}.
   \label{eqn:main_Ham}
\end{align}
{We use periodic boundary condition (PBC) throughout the paper. However, the effects of open boundary conditions (OBC), particularly in the context of Hilbert space fragmentation, are presented in Appendix~\ref{sec:obc}. Here $\sigma^x_i$ is the Pauli $x$ operator at site $i$, and the Hamiltonians $H^{\mathcal{N}(k)}$ described above each represents a hierarchical level of kinetic constraints that govern the dynamics at site $i$, as determined by the operator $\mathcal{N}_i^{(k)}$. Specifically, the  
\(\mathcal{N}_i^{(k)}\)'s are operators defined on four adjacent sites of $i$ (left and right nearest neighbors $i-1, i+1$ and left and right next-nearest neighbors $i-2,i+2$) that ensure the evolution criterion is met: {\it the quantum state at site $i$ will flip provided the sum of the population of its four adjacent sites equals $k$}. Mathematically, we can define them as follows: 

\begin{eqnarray} 
\mathcal{N}_i^{(0)} &=& {n}_{i-2} n_{i-1}n_{i+1}n_{i+2}, \nonumber \\ 
\mathcal{N}_i^{(1)} &=& \bar{n}_{i-2} n_{i-1}n_{i+1}n_{i+2}+n_{i-2}\bar{n}_{i-1} n_{i+1}n_{i+2} \nonumber \\ 
    &+& n_{i-2}n_{i-1}\bar{n}_{i+1}n_{i+2} +n_{i-2}n_{i-1}n_{i+1}\bar{n}_{i+2}, \nonumber\\
\mathcal{N}_i^{(2)} &=& \bar{n}_{i-2}\bar{n}_{i-1}n_{i+1}n_{i+2}+\bar{n}_{i-2}n_{i-1}\bar{n}_{i+1}n_{i+2} \nonumber \\
    &+&\bar{n}_{i-2}n_{i-1}n_{i+1}\bar{n}_{i+2} + n_{i-2}\bar{n}_{i-1}\bar{n}_{i+1}n_{i+2} \nonumber \\&+& n_{i-2}\bar{n}_{i-1}n_{i+1}\bar{n}_{i+2} + n_{i-2}n_{i-1}\bar{n}_{i+1}\bar{n}_{i+2}, \nonumber \\ 
   \mathcal{N}_i^{(3)} &=& \bar{n}_{i-2}\bar{n}_{i-1}\bar{n}_{i+1}n_{i+2} + \bar{n}_{i-2}\bar{n}_{i-1}n_{i+1}\bar{n}_{i+2} \nonumber \\ 
   &+& \bar{n}_{i-2}n_{i-1}\bar{n}_{i+1}\bar{n}_{i+2} + n_{i-2}\bar{n}_{i-1}\bar{n}_{i+1}\bar{n}_{i+2},
   \label{eqn:N_Ham}
\end{eqnarray} 
with $n_k=|0\rangle \langle 0|_k$ and $\bar{n}_k= |1\rangle \langle 1|_k$. From the above expression we can see  \( H^{\mathcal{N}(0)} \) is unconstrained PPXPP model that  unlike the conventional PPXPP model, does not include the Rydberg constraint, which restricts the basis to have configurations where any site in state $|1\rangle$ is always surrounded by at least two zeros  on both its left and right, i.e., of the form $|\dots 00100 \dots \rangle$. Moreover, Eq.~(\ref{eqn:N_Ham}) explicitly shows that each   $H^{\mathcal{N}(k)}$ consists of terms involving asymmetric projectors. 

\noindent
In our work, along with the quantum properties of the Hamiltonian defined in Eq.~(\ref{eqn:main_Ham}) and individual \( H^{\mathcal{N}(k)} \)'s, we also consider quantum Hamiltonians that arise from various combinations of different \( H^{\mathcal{N}(k)} \). In what follows, we briefly introduce them for future reference.

\noindent
(a) {\it Perturbed PPXPP model}: The first case we consider in our work is a combination of \( H^{\mathcal{N}(0)} \) (unconstrained PPXPP) and \( H^{\mathcal{N}(k)} \), where $k=1$ and 2. The resulting Hamiltonian is denoted by 
\begin{equation}
H^{\text{Pert}}_{\text{PPXPP}}(k) = H^{\mathcal{N}(0)}+\delta \cdot H^{\mathcal{N}(k)}=\sum_{i=1}^L \sigma^x_i  (\mathcal{N}_i^{(0)}+\delta \cdot \mathcal{N}_i^{(k)}),
\end{equation}
where $\delta$ is a constant. In Sec.~\ref{sec:quantum scars}, we discuss how this model lead to the remarkable coexistence of quantum many-body scars and Hilbert space fragmentation within it. \\ \noindent
(b) {\it Quantum Game of Life}: One notable case is the combination of \( H^{\mathcal{N}(2)} \) and \( H^{\mathcal{N}(3)} \), which corresponds to the well-known Quantum Game of Life (QGL) model, defined as
\begin{equation}
H^{\text{QGL}} =H^{\mathcal{N}(2)}+H^{\mathcal{N}(3)}= \sum_{i=1}^L \sigma^x_i  (\mathcal{N}_i^{(2)}+\mathcal{N}_i^{(3)}).
\end{equation}
This model being an example of celebrated cellular automata,  extends the classical Game of Life introduced by John Conway in 1973~\cite{gardner1970life}, which is known to be Turing-complete~\cite{schulman1978life,rendell2002collision}. The quantum extension explores the potential for simulating universal quantum computation~\cite{bleh2012qgol,ney2022life}. In our case, we conduct an in-depth study of the model by examining the effects of the components \( H^{\mathcal{N}(2)} \) and \( H^{\mathcal{N}(3)} \) separately and their interplay with \(H^{\text{QGL}} \).

In the literature, there have been previous proposals for implementation of  quantum kinetically constrained models that belong to the paradigm of quantum cellular automata~\cite{Prosen2021_RCA_IRF} with Rydberg atoms~\cite{Wintermantel_2020}. The main idea is to use programmable multifrequency excitation and depumping of Rydberg states to create conditional interactions that are similar to classical cellular automata.  Importantly, the essential ingredient in such realizations is not a specific blockade-induced Hamiltonian, but the ability to implement local, rule-dependent evolution conditioned on the surrounding population. 
Furthermore, recent experimental advances have demonstrated QCA-like dynamics in dual-species Rydberg processors
where programmable constraint rules and conditional dynamics can be engineered at the hardware level (see \cite{white2026}). These developments highlight the growing capability to realize and probe rule-based quantum evolution in controlled settings.  In our case, we expect that the rule-based KCMs studied here, including the quantum Game of Life, can be implemented using similar programmable Rydberg platforms by encoding the update rules through conditional excitation and depumping protocols.

In the forthcoming sections, we demonstrate that although these models are structurally very similar, the presence of different kinetic constraints leads to significantly different quantum properties.

\section{Hamiltonian-Induced Hilbert Space Structure}
\label{Sec:Hamiltonian-Induced Hilbert Space Structure}
The presence of kinetic constraints in all the quantum many-body Hamiltonians introduced above affects the Hilbert space structure directly. Here, we provide a detailed discussion about them that will be instrumental in understanding the behavior of quantum properties we explore in the forthcoming section.

\subsection{Effective Hilbert space dimension}
One such direct consequence of the kinetic constraints is the annihilation of certain basis states. As a result, the system is unable to fully explore the entire Hilbert space, since some regions become dynamically inaccessible. The effective Hilbert space dimension (i.e. excluding the trivially annihilated states) in this case depends on the type of constraints and the combination of the Hamiltonians considered. In Table~\ref{table_dimension}, we list them down for different  \( H^{\mathcal{N}(k)}\)'s as well as \(H^{\text{QGL}}\).

\begin{table}[h]
\centering
\begin{tabular}{|c|c|c|c|c|c|c|c|c|}
\hline
$L$ & 5 & 6 & 7 & 8 & 9 & 10 & 11 & 12 \\
\hline
$d^{0}_L$ & 6 & 10 & 29 & 77 & 175 & 376 & 793 & 1682 \\
\hline
$d^{1}_L$ & 15 & 30 & 84 & 184 & 396 & 835 & 1716 & 3530 \\
\hline
$d^{2}_L$ & 20 & 50 & 112 & 234 & 492 & 992 & 2002 & 4018 \\
\hline
$d^{3}_L$ & 15 & 30 & 84 & 184 & 396 & 835 & 1716 & 3530 \\
\hline
$d^{\mathrm{QGL}}_L$ & 25 & 56 & 119 & 246 & 501 & 1007 & 2024 & 4064 \\
\hline
$2^L$ & 32 & 64 & 128 & 256 & 512 & 1024 & 2048 & 4096 \\
\hline
\end{tabular}
\caption{Scaling of the effective  Hilbert space dimensions ($d$) with system size $L$ of kinetically constrained Hamiltonians $H^{\mathcal{N}(k)}$, for $k = 0,1,2,3$ and  $H^{\text{QGL}}$} 
\label{table_dimension}
\end{table}

\subsection{Symmetry resolutions} 
One of the crucial steps of our analysis is identifying and resolving all possible symmetries present in the set of Hamiltonians described above. Such symmetry reductions help us access larger system sizes and reveal the true behavior of the spectral characteristics. Below, we list all of them.\\

\noindent
(a) {\bf \it Translational symmetry}: With  periodic boundary conditions, all the Hamiltonians $\{H^{\mathcal{N}(k)}, H^{\text{QGL}}, H^{\text{Pert}}_{\text{PPXPP}}(k)\}$ exhibit translation symmetry. Let $\hat{\mathcal{T}}$ be the translation operator which shifts the $i$'th site to the $(i+1)$'th site. This operator commutes with all the Hamiltonians. The eigenvalue of this operator is $e^{ik}$, where $k$ is the momentum of that eigenstate. In our case, we primarily focus on the translationally invariant zero-momentum ($k=0$) subspace.  However, as shown in Appendix \ref{sec_k_other}, choosing subspaces other than $k=0$ does not significantly affect the results, provided that the Hilbert space dimension remains comparable.\\

\noindent 
(b) {\bf \it Inversion symmetry}: All the Hamiltonians commute with the inversion operator, which performs an inversion about the midpoint of the spin chain. If we denote the inversion operator as \(\hat{\mathcal{I}}\), its effect on site \(i\) can be expressed as: \(i \rightarrow L - i + 1\). When the operator is applied twice, it returns the state to its original configuration. Consequently, \(\hat{\mathcal{I}}\) has eigenvalues of \(\pm 1\). This leads to the formation of two sectors: the inversion-symmetric sector with \(\mathcal{I} = +1\) and the inversion-antisymmetric sector with \(\mathcal{I} = -1\).\\
\noindent 
 (c) {\bf \it Chiral symmetry}: We also observe that our set of Hamiltonians anti-commutes with a Hermitian unitary operator, denoted as \(\hat{\mathcal{C}}\), which is referred to as the chiral operator \(\hat{\mathcal{C}} = \prod_i \sigma^z_i\). This symmetry does not give rise to any additional conserved quantities. However, as a result, the energy spectrum becomes symmetric around the value \(E = 0\) and leads to a significant number of zero-energy eigenstates.\\
\noindent 
 (d) {\bf \it Spin-flip or global $\mathbb{Z}_2$ symmetry}:
The final symmetry we will discuss here is the spin-flip or also known as global $\mathbb{Z}_2$ symmetry exhibited by \( H^{\mathcal{N}(2)} \). This Hamiltonian commutes with the spin-flip operator \(\hat{\kappa} = \prod_i \sigma^x_i\). The operator has eigenvalues of \(\kappa = \pm 1\), leading to symmetric and antisymmetric spin-flip sectors. For the rest of the Hamiltonians, the effect of this local spin-flip operation results in
\begin{eqnarray}
\left(\prod_i  \sigma^x_i\right) H^{\mathcal{N}(k)} \left(\prod_i \sigma^x_i\right)  =H^{\mathcal{N}(4-k)}. 
\end{eqnarray}
For instance,  it maps $H^{\mathcal{N}(1)}$ to $H^{\mathcal{N}(3)}$ and vice versa. \\
\noindent

\subsection{Hilbert Space Fragmentation} 
\label{sec:Hil_frag}
Symmetries typically divide the Hilbert space into different subspaces that do not interact with each other and are characterized by conserved quantum numbers. However, due to the presence of kinetic constraints in the system, the Hilbert space may get ``fragmented" into exponentially many different additional subspaces. This phenomenon is commonly known as Hilbert space fragmentation~\cite{Moudgalya_2022,Mukherjee2021,Anwesha2023,Francica2023}. The term was first coined in~\cite{Sala2020}. The number of subspaces due to symmetry normally stays constant with the number of lattice sites or at most grows polynomially. However, in the case of strong HSF, the behavior is opposite, and the number of subspaces can grow exponentially with system size $L$. The quantum states $|\psi\rangle$ in the dynamically disconnected spaces, commonly referred to as ``Krylov subspaces'', remain confined to the same subspace under the action of the Hamiltonian. In other words, the full Hilbert space can be characterized by
\begin{figure}
  \includegraphics[width=0.48\textwidth]{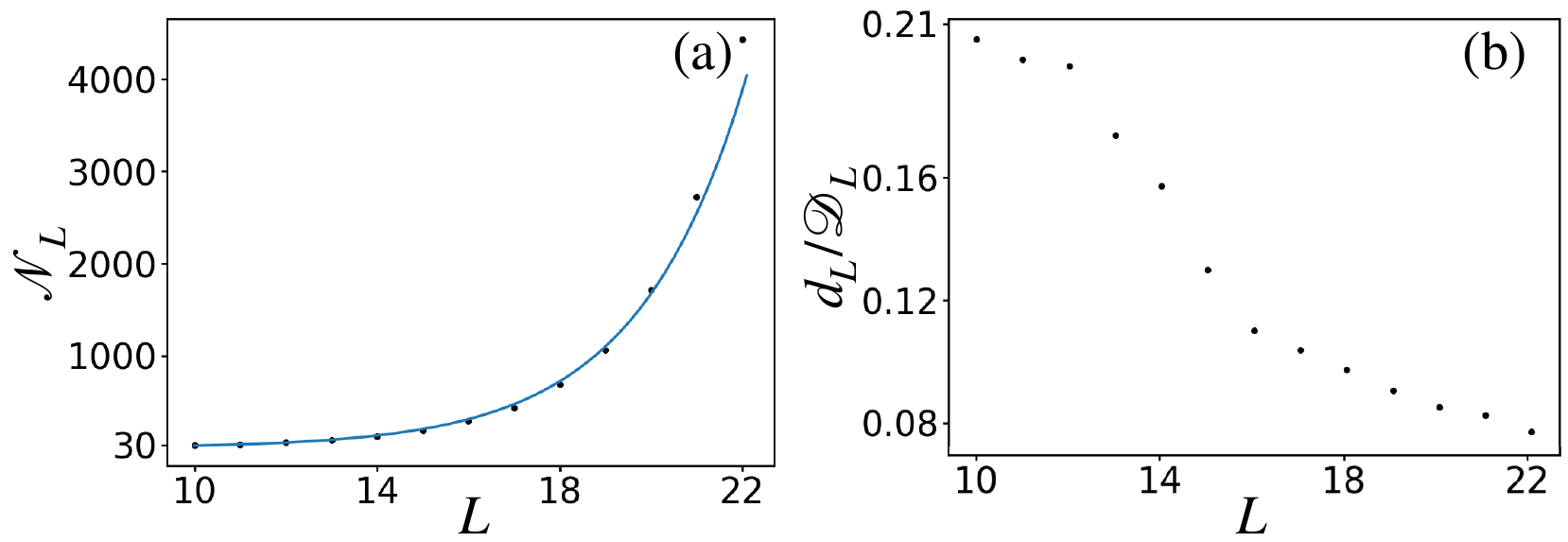}
  \caption{Krylov subspaces  and their scaling in fragmented model \(H^{\mathcal{N}(1)}\).   In (a) and (b), we illustrate how the number of Krylov subspaces, \(\mathcal{N}_L\), and the ratio \(d_L/\mathcal{D}_L\) behaves with system size \(L\), respectively. The maximum system size considered is \(L = 22\). Both figures clearly indicate the presence of strong Hilbert space fragmentation (HSF). In contrast, \(H^{\mathcal{N}(2)}\)  displays a slower decay of \(d_L/\mathcal{D}_L\) with increasing \(L\). }
  \label{Fig_Strong_HSL}
\end{figure}

\begin{equation}
    \mathcal{H}= \bigoplus_{i=1}^{\mathcal{N}} \mathcal{K}_i\ , \ \ \mathcal{K}_i = \{|\psi_i\rangle, H |\psi_i\rangle, H^2 |\psi_i\rangle, \cdots\} \ .
\end{equation}
Here, $\mathcal{N}$ is the number of Krylov subspaces. For system size $L$, and a symmetry-resolved sector with Hilbert space dimension $\mathcal{D}_L$, the dimension of the largest Krylov subspace is denoted by $ d_L$. If $d_L/\mathcal{D}_L \rightarrow 0$ as $L \rightarrow \infty$, this is called strong Hilbert space fragmentation. If $d_L/\mathcal{D}_L \rightarrow 1$ as $L \rightarrow \infty$, it is called weak Hilbert space fragmentation.

We observe that, in the symmetry-resolved subspace ($k=0$) of $H^{\mathcal{N}(0)}$, Hilbert space fragmentation is present. 
However, this fragmentation is predominantly trivial and is dominated by Krylov subspaces of dimension one. 
These correspond to basis states that are completely frozen by the kinetic constraint and are annihilated by the Hamiltonian, and therefore do not participate in any nontrivial dynamics. Interestingly, as is known from the literature, the PPXPP model with the Rydberg constraint does not exhibit any fragmentation~\cite{Kerschbaumer_2025}, which can be reverified from our analysis. In Fig.~\ref{fig_fragH0} in Appendix~\ref{sec:fragH0}, for $L=7$, we present a schematic of the fragmented subspaces obtained for both $H^{\mathcal{N}(0)}$ and the PPXPP model.  We can see the additional condition coming from the Rydberg constraint eliminates all but the largest-dimensional Krylov space of $H^{\mathcal{N}(0)}$, thereby removing the fragmentation in the PPXPP model.

\begin{figure}
  \includegraphics[width=.5\textwidth]{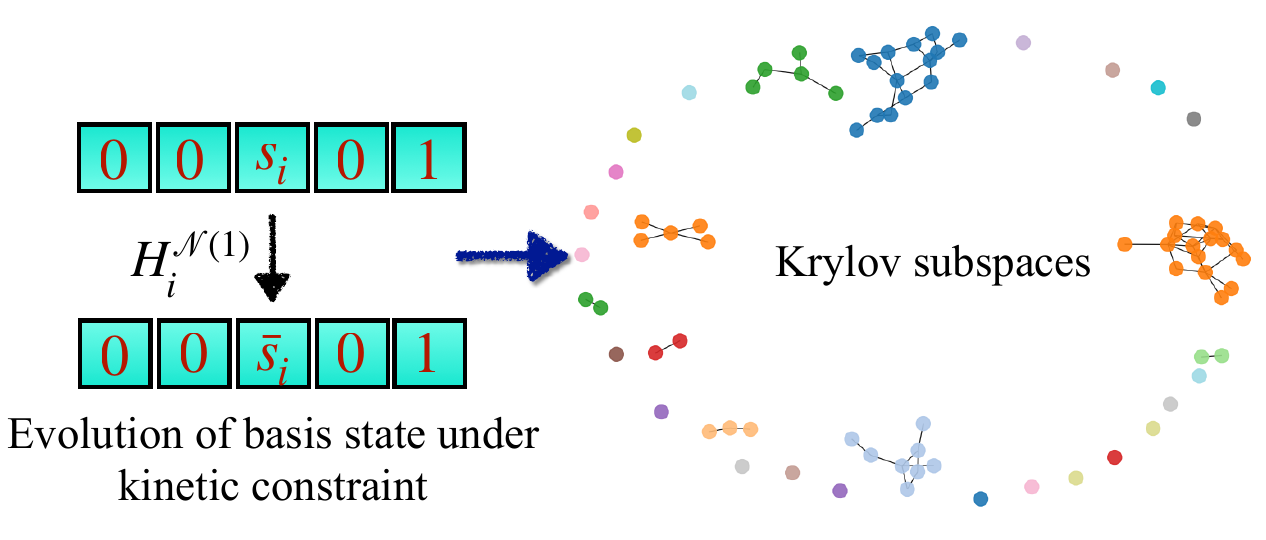}
  \caption{Graphical representation of Hilbert space fragmentation due to kinetic constraints in the Hamiltonian $H^{\mathcal{N}(1)}$ [Eq.~(\ref{eqn:main_Ham})], in zero momentum inversion symmetric sector. The kinetic constraint allows the quantum state at site $i$ ($s_i$) to evolve only when the combined occupation of its surrounding neighbors ($i-2, i-1, i+1, i+2$) adds up to one, as illustrated on the left. On the right, different colors denote dynamically disconnected Krylov sectors for a system of size $L=10$. Each circle represents a basis state, and an edge indicates that the state can evolve into the connected state under the action of the Hamiltonian. }
  \label{fig:hsf_sche}
\end{figure}

\begin{table}[htbp]
\centering
\small
\resizebox{\columnwidth}{!}{%
\begin{tabular}{|c|c|c|c|c|c|c|c|c|c|c|c|c|c|}
\hline
$L$ & 10 & 11 & 12 & 13 & 14 & 15 & 16 & 17 & 18 & 19 & 20 & 21 & 22 \\
\hline
$\mathcal{N}^{1}_L$ & 30 & 39 & 61 & 82 & 127 & 185 & 288 & 431 & 690 & 1063 & 1717 & 2722 & 4434 \\
\hline
$d^{1}_L$ & 16 & 25 & 44 & 66 & 108 & 159 & 248 & 427 & 749 & 1295 & 2298 & 4202 & 7472 \\
\hline
$\mathcal{D}^{1}_L$ & 78 & 126 & 224 & 380 & 687 & 1224 & 2250 & 4112 & 7685 & 14310 & 27012 & 50964 & 96909 \\
\hline
$\mathcal{N}^{2}_L$ & 8 & 9 & 12 & 12 & 15 & 17 & 20 & 22 & 27 & 31 & 38 & 43 & 54 \\
\hline
$d^{2}_L$ & 14 & 21 & 37 & 58 & 102 & 177 & 312 & 564 & 1003 & 1869 & 3360 & 6334 & 11488 \\
\hline
$\mathcal{D}^{2}_L$ & 44 & 63 & 122 & 190 & 362 & 612 & 1162 & 2056 & 3914 & 7155 & 13648 & 25482 & 48734 \\
\hline
\end{tabular}%
}
\caption{The relevant dimensions of Hilbert spaces and the counts of dynamically disconnected sectors for different system sizes \( L \) shown for \( H^{\mathcal{N}(1)} \) and \( H^{\mathcal{N}(2)} \). In this context, we denote the number of Krylov subspaces as \( \mathcal{N}_L \),  the dimension of the Hilbert space after symmetry resolution as \( \mathcal{D}_L \), and the dimension of the largest Krylov subspace as \( d_L \).}
\label{table_dimension_frag}
\end{table}

We further report that \( H^{\mathcal{N}(1)} \) exhibits HSF with the number of fragmented subspaces increasing exponentially with the size of the system \( L \), as shown in Fig.~\ref{Fig_Strong_HSL}(a). Additionally, in this case, the ratio \( d_L^1/\mathcal{D}_L^1 \) approaches zero as \( L \) approaches infinity, indicating that \( H^{\mathcal{N}(1)} \) demonstrates a strong form of Hilbert space fragmentation (HSF), as depicted in Fig.~\ref{Fig_Strong_HSL}(b).  A schematic representation of the fragmented subspaces of $H^{\mathcal{N}(1)}$ in zero momentum inversion symmetric sector for this model is shown in Fig.~\ref{fig:hsf_sche} for $L=10$. Moreover, unlike $H^{\mathcal{N}(0)}$, the Krylov subspaces in this case consist of more than one basis state and are not trivially annihilated solely due to the kinetic constraint. Hence, it is thus important to distinguish between \emph{trivial} and \emph{emergent} Hilbert space fragmentation. 

In contrast, for $H^{\mathcal{N}(2)}$, although we observed HSF,  the number of fragmented subspaces increases polynomially  with the size of the system. The fragmentation structure of \( H^{\mathcal{N}(3)} \) remains exactly the same as \( H^{\mathcal{N}(1)} \) since both are related by the spin-flip symmetry.  
Notably, while the individual Hamiltonians \( H^{\mathcal{N}(2)} \) and \( H^{\mathcal{N}(3)} \) display distinct behaviors, their sum, in other words,  \( H^{\text{QGL}} \), shows no fragmentation at all. Apart from the trivial states that are annihilated by the Hamiltonian, the entire Hilbert space is instead highly connected. Similar behavior can be observed for $H^{\text{Tot}}$.

In  Table~\ref{table_dimension_frag}, we show the number of Krylov subspaces $(\mathcal{N}_L)$, the dimension of the Hilbert space after the resolution of the symmetries $(\mathcal{D}_L)$, and the dimension of the largest subspace $(d_L)$ among them,  for $H^{\mathcal{N}(1)}$, $H^{\mathcal{N}(2)}$ with increasing $L$. Here, we have also counted the trivial annihilated states. For  $H^{\text{Pert}}_{\text{PPXPP}}$, which is a combination of the Hamiltonians \(H^{\mathcal{N}(0)}\) and \(H^{\mathcal{N}(1)}\) (\(H^{\mathcal{N}(2)}\)), the HSF persists with  richer structure of   fragmented subspaces than that obtained for \(H^{\mathcal{N}(1)}\) (\(H^{\mathcal{N}(2)}\))  (as can be seen from Table \ref{Table:H_ppxpp_purturbed}). We additionally present the behavior of $d_L/\mathcal{D}_L$ with $L$ for the remaining fragmented KCMs, in Fig.~\ref{fig:HSF_scaling} , Here we can see that the Hamiltonian $H^{\mathcal{N}(0)}$ exhibits strong fragmentation. But it contains 
many Krylov subspaces of dimension $1$, i.e. frozen states, leading to a strongly suppressed
$d_L/\mathcal{D}_L$. $H^{\mathcal{N}(2)}$ shows a clear even-odd effect in its scaling with $L$.
All three models shown in (a), (b) and (c) in Fig.~\ref{fig:HSF_scaling} show that $d_L/\mathcal{D}_L$ decreases as L increases. However, the model $H_{\text{PPXPP}}^{\text{Pert}}(k=2)$ (shown in (d)) shows that the ratio $d_l/\mathcal{D}_L$ is saturating around 0.5 with increase in $L$, giving a qualitatively different behavior than others.\\

\begin{table}[htbp]
\centering

\subtable[\(k=1\)]{
\resizebox{0.87\columnwidth}{!}{%
\begin{tabular}{|c|c|c|c|c|c|c|c|c|c|c|c|c|c|}
\hline
$L$ & 10 & 11 & 12 & 13 & 14 & 15 & 16 & 17 & 18 & 19 & 20 & 21 &22 \\
\hline
$\mathcal{N}^{01}_L$ & 24 & 31 & 49 & 66 & 103 & 148 & 231 & 343 & 548 & 838 & 1348 & 2117 & 3425 \\
\hline
$d^{01}_L$ & 33 & 51 & 84 & 132 & 220 & 361 & 609 & 1019 & 1749 & 2978 & 5345 & 9793 & 18023 \\
\hline
$\mathcal{D}^{01}_L$ & 78 & 126 & 224 & 380 & 687 & 1224 & 2250 & 4112 & 7685 & 14310 & 27012 & 50964 & 96909 \\
\hline
\end{tabular}
}}

\hfill

\subtable[\(k=2\)]{
\resizebox{0.87\columnwidth}{!}{%
\begin{tabular}{|c|c|c|c|c|c|c|c|c|c|c|c|c|c|}
\hline
$L$ & 10 & 11 & 12 & 13 & 14 & 15 & 16 & 17 & 18  & 19 & 20 & 21 & 22 \\
\hline
$\mathcal{N}^{02}_L$ & 5 & 5 & 8 & 7 & 8 & 9 & 11 & 12 & 14 & 16 & 20 & 22 & 27 \\
\hline
$d^{02}_L$ & 38 & 64 & 113 & 191 & 341 & 619 & 1127 & 2056 & 3840 & 7165 & 13532 & 25516 & 48458  \\
\hline
$\mathcal{D}^{02}_L$ & 78 & 126 & 224 & 380 & 687 & 1224 & 2250 & 4112 & 7685 & 14310 & 27012 & 50964 & 96909 \\
\hline
\end{tabular}
}}

\caption{Relevant Hilbert space dimensions and Krylov sector counts for the perturbed Hamiltonian
\( H^{\text{Pert}}_{\text{PPXPP}}(k) = H^{\mathcal{N}(0)} + \delta\, H^{\mathcal{N}(k)} \).
For a system of size \(L\), \(\mathcal{N}^{0k}_L\) denotes the number of Krylov subspaces,
\(d^{0k}_L\) the dimension of the largest Krylov subspace, and
\(\mathcal{D}^{0k}_L\) the symmetry-resolved Hilbert space dimension.}
\label{Table:H_ppxpp_purturbed}
\end{table}

Hence, our study reveals that the nature of Hilbert-space fragmentation can vary qualitatively across this family of models, ranging from strong to weak fragmentation and ultimately to its complete absence. This behavior depends sensitively on the underlying population-dependent rules and, to the best of our knowledge, has not been explicitly demonstrated within a single family of kinetically constrained models. Consequently, predicting such behavior \textit{a priori} remains highly nontrivial without detailed numerical analysis.


\begin{figure}
    \centering
    \includegraphics[width=0.5\textwidth]{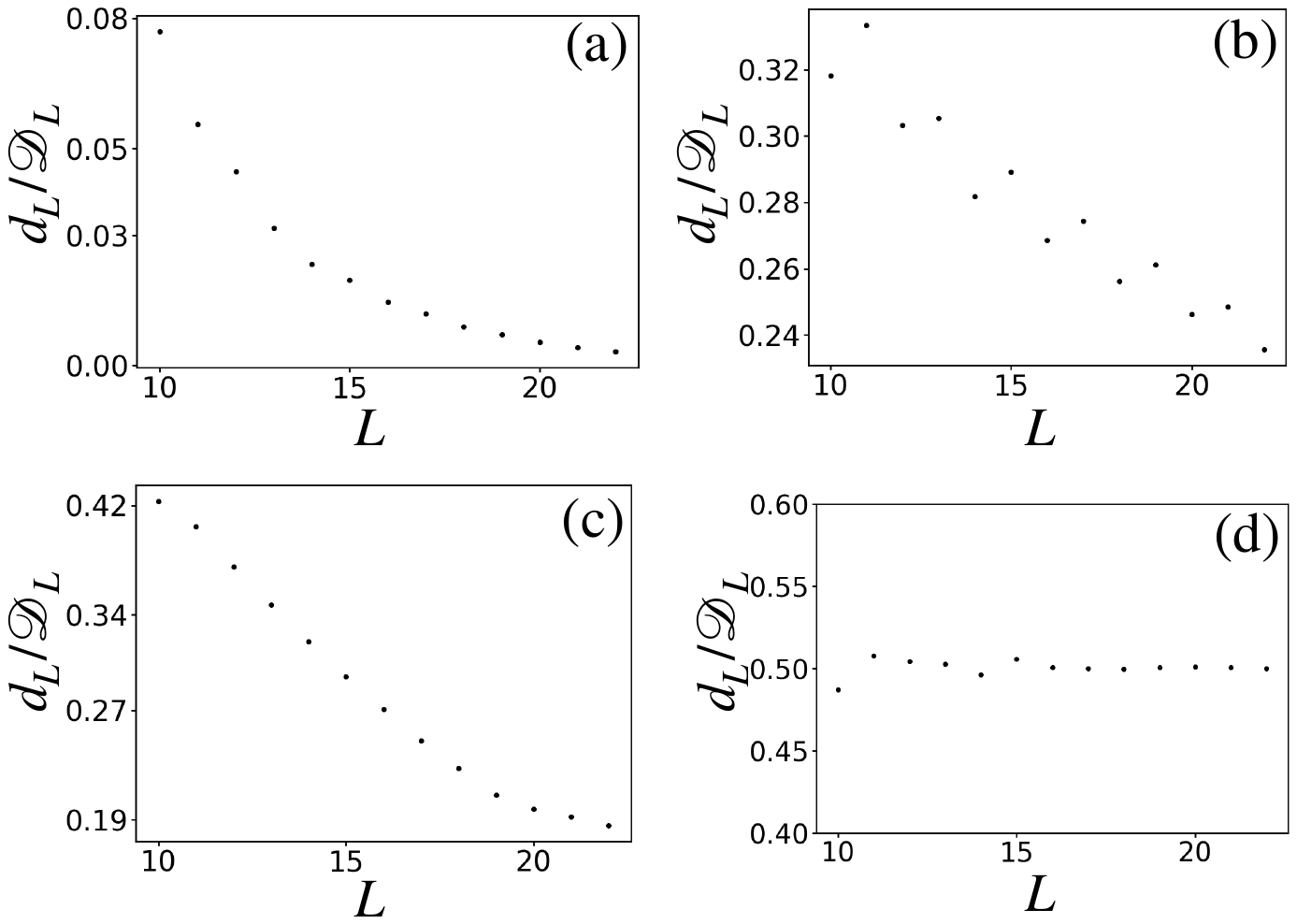} 
    \caption{Scaling of biggest Krylov subspace in fragmented models.
Shown is the ratio $d_L/\mathcal{D}_L$ as a function of system size $L$ for
(a) $H^{\mathcal{N}(0)}$,
(b) $H^{\mathcal{N}(2)}$,
(c) $H^{\mathrm{Pert}}_{\mathrm{PPXPP}}(k=1)$, and
(d) $H^{\mathrm{Pert}}_{\mathrm{PPXPP}}(k=2)$.
}
 
    \label{fig:HSF_scaling}
\end{figure}

\section{Quantum complexity study} 
\label{Sec:quantum complexity}
As mentioned earlier, in our work, we carried out a thorough analysis of the interplay of  quantum state preparation complexity measures, namely, entanglement and nonstabilizerness, and markers of dynamical quantum complexity such as quantum chaos. In the following, we will begin the discussion with the behavior of entanglement in these models and conduct a comparative study with other forms of complexities.
\subsection{Entanglement}
For the entanglement analysis, we compute the half-chain von Neumann entanglement entropy (EE) between two equal halves $A$ and $B$ of the system. it is defined by:
\begin{align}
S_{L/2} &=- \text{Tr} \Big[ \rho_{\text{A}} \log_2 (\rho_{\text{A}}) \Big]= - \sum_i \lambda_i \log_2 \lambda_i,
\end{align}
where $\rho_A= \text{Tr}_{\text{B}} (\rho_{\text{AB}})$ is the reduced density matrix of subsystem $A$ and $\lambda_i$s are its eigenvalues.
\begin{figure}
  \includegraphics[width=.48\textwidth]{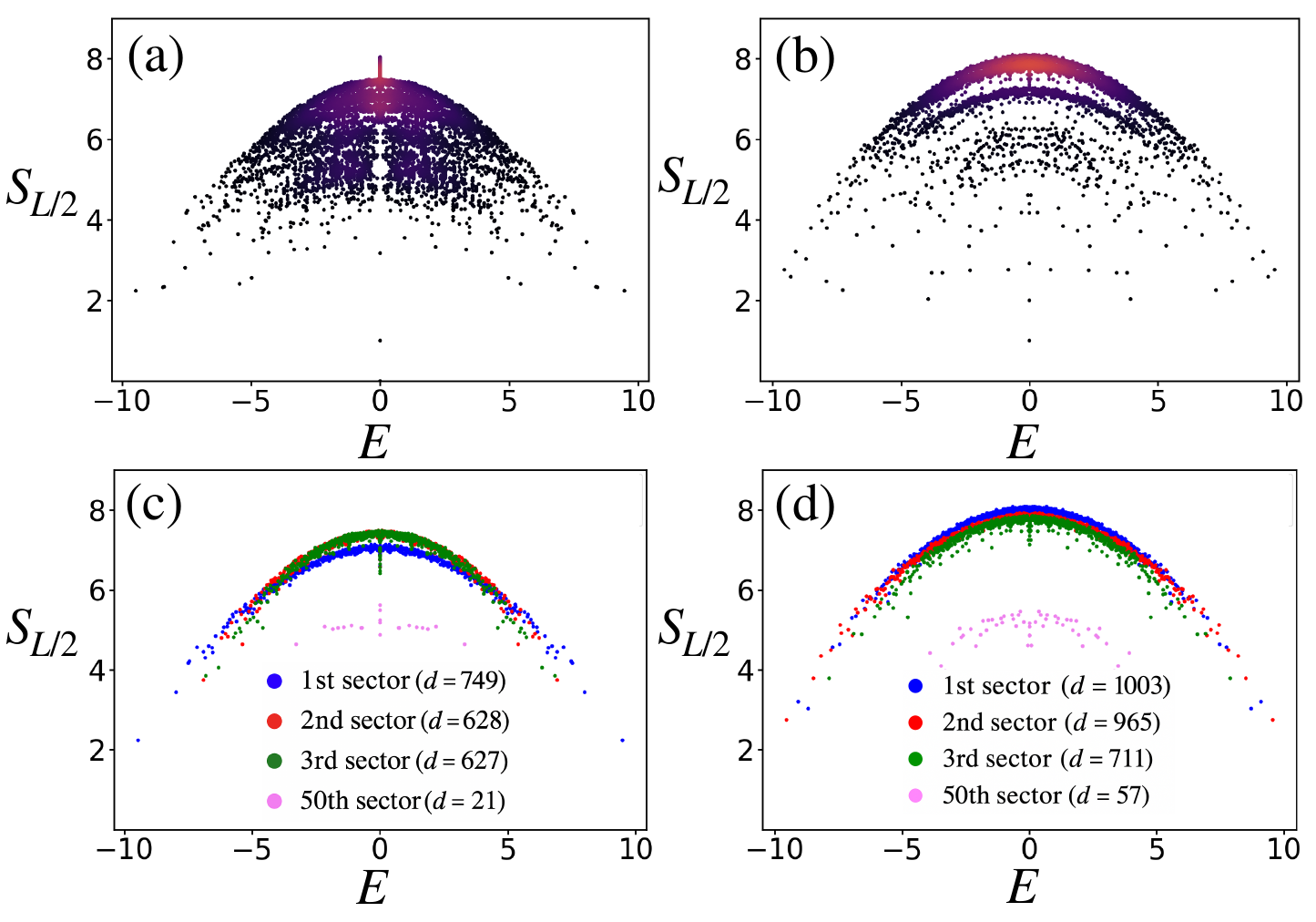}
  \caption{Behavior of  entanglement entropy for the eigenstates of kinetically  constrained quantum many-body models \( H^{\mathcal{N}(1)} \) and \( H^{\mathcal{N}(2)} \). In (a), we present the behavior of half-chain entanglement entropy (\( S_{L/2} \)) for the part of the spectrum of  \( H^{\mathcal{N}(1)} \) belonging to the zero momentum and inversion symmetric sector for \( L = 18 \). For \(H^{\mathcal{N}(2)}\), we obtained a similar behavior in (b) for the subspace that along with the translation symmetry (\(k=0\) subspace) and inversion symmetry, has an additional spin-flip symmetry. In (c), we present the same plot as shown in (a), but highlighting the entanglement behavior of four specific Krylov subspaces using different colors. The three largest subspaces are of dimensions 749, 628, and 627, while the 50th largest subhas a dimension of only 21. Notably, for \( H^{\mathcal{N}(1)} \), the subspace with the largest dimension does not correspond to the highest entanglement entropy; instead, the highest entanglement comes from the second and third largest subspaces. Conversely, for \( H^{\mathcal{N}(2)} \), as shown in (d), the largest Hilbert space, with dimension \( d_L = 1003 \) (marked in blue), produces the highest entanglement entropy. The next two largest fragmented subspaces have dimensions of 965 and 711, marked in red and green, respectively.}
  \label{fig:ent_h1_h2}
\end{figure}
The behavior of EE becomes crucial in unveiling  different layers of physics associated with the models. Fig.~\ref{fig:ent_h1_h2}(a) shows that for $H^{\mathcal{N}(1)}$, a larger fraction of the eigenstates, even those near the middle of the energy spectrum, significantly deviate from the characteristic ``arch"-like shape typically associated with thermal entanglement. Interestingly,  despite $H^{\mathcal{N}(2)}$ having an additional symmetry than $H^{\mathcal{N}(1)}$, Fig.~\ref{fig:ent_h1_h2}(b) shows it allows a larger fraction of bulk eigenstates to approach the arch value. This is a  consequence of less stringent kinetic constraints imposed by $H^{\mathcal{N}(2)}$  that allows the system to explore a larger effective Hilbert space than $H^{\mathcal{N}(1)}$ (see Table~\ref{table_dimension}). However, in both cases, the states appearing below the arch are a direct consequence of the Hilbert space fragmentation. This becomes clear when we examine the entanglement characteristics of the fragmented subspaces separately, as shown in Figs.~\ref{fig:ent_h1_h2}(c) and (d). In this case, we present the entanglement behavior of each HSF subspace using distinct colours, resulting in the appearance of multiple arches, each corresponding to a fragmented subspace.  For example, for $H^{\mathcal{N}(1)}$, the eigenstates of the second (dimension = 628, marked by red) and third (dimension = 627, marked by green) largest fragmented subspaces form arches that lie almost on top of each other. Whereas, the largest subspace's (dimension=749, marked by blue) entanglement remains below that.

\begin{figure}
  \includegraphics[width=.42\textwidth]{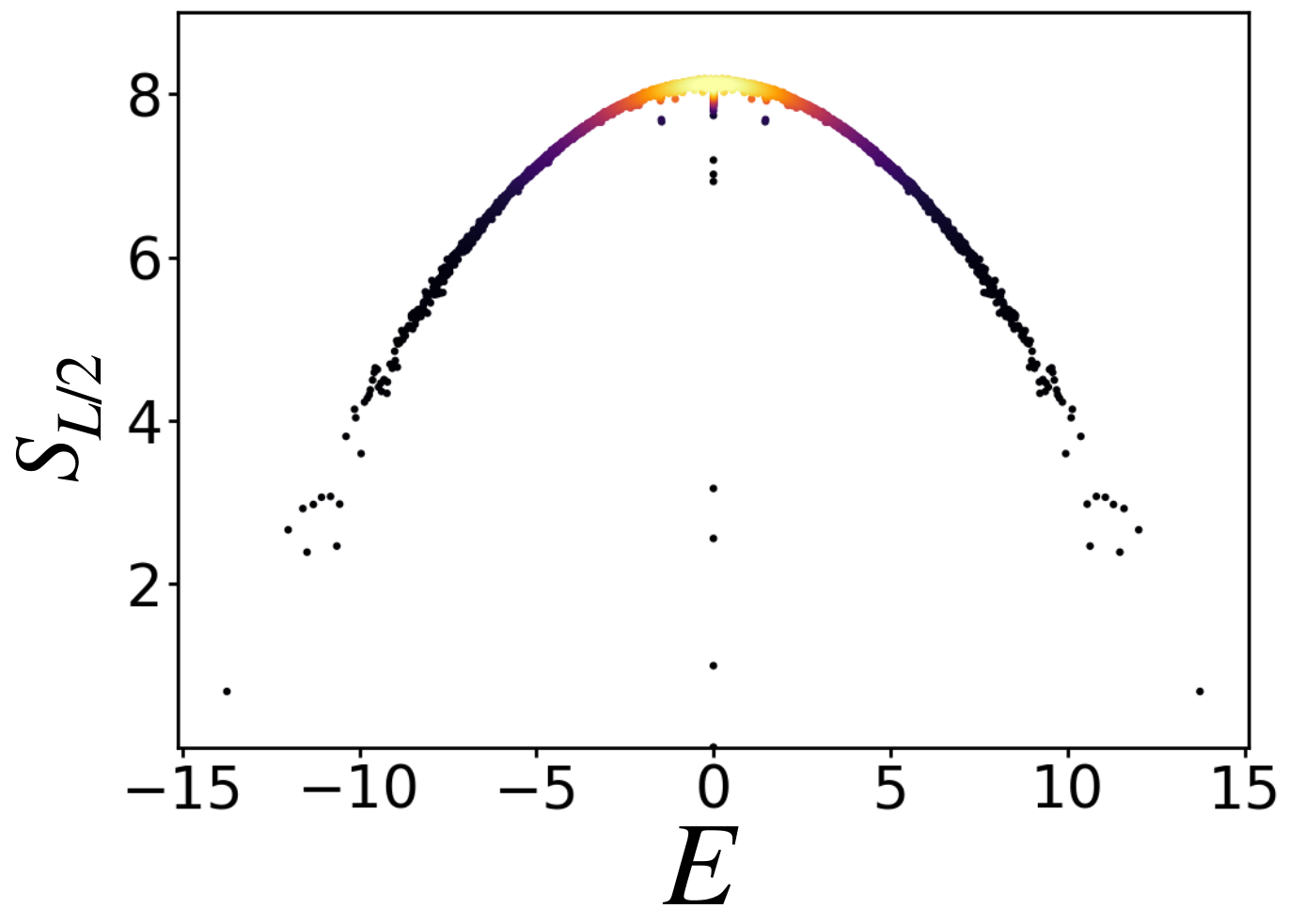}
  \caption{Behavior of entanglement entropy  for the eigenstates of the models QGL. Here we plot the behavior of half-chain entanglement entropy ($S_{L/2}$) of the eigenstates of $H^{\text{QGL}}$ in the zero momentum and inversion symmetric sector obtained for $L=18$. The behavior shows smooth variations of bipartite entanglement with energy.}
  \label{fig:ent_qgl}
\end{figure}

In the case of $H^{\mathcal{N}(2)}$, the dimensions of the three largest fragmented subspaces are 1003, 965, and 711, and the corresponding arches are marked in blue, red, and green, respectively. We can see from Fig.~\ref{fig:ent_h1_h2}(d) that  in this case, the arches follow the same ordering as their Hilbert space sizes.   It is observed that in both the cases,   $H^{\mathcal{N}(1)}$  and  $H^{\mathcal{N}(2)}$,  the Krylov subspaces with lower dimensions fail to exhibit thermal behavior since they occupy a very small portion of the full Hilbert space. Furthermore, these smaller subspaces tend to have low entanglement entropy, contributing to the variation along the y-axis in Figures \ref{fig:ent_h1_h2}(a) and (b). Therefore, our analysis characterizes these fragmented subspaces according to their entanglement-generating capabilities, revealing that the largest dimension space does not always correspond to the one that generates the highest amount of entanglement entropy.  Thereafter, we observe that as the HSF effect diminishes in the case of QGL, the symmetry-resolved subspace shows a consistent entanglement entropy (EE) profile as shown in Fig.~\ref{fig:ent_qgl}. This results in a well-shaped arch, with very few eigenstates away from it. 

\begin{figure}
  \includegraphics[width=.48\textwidth]{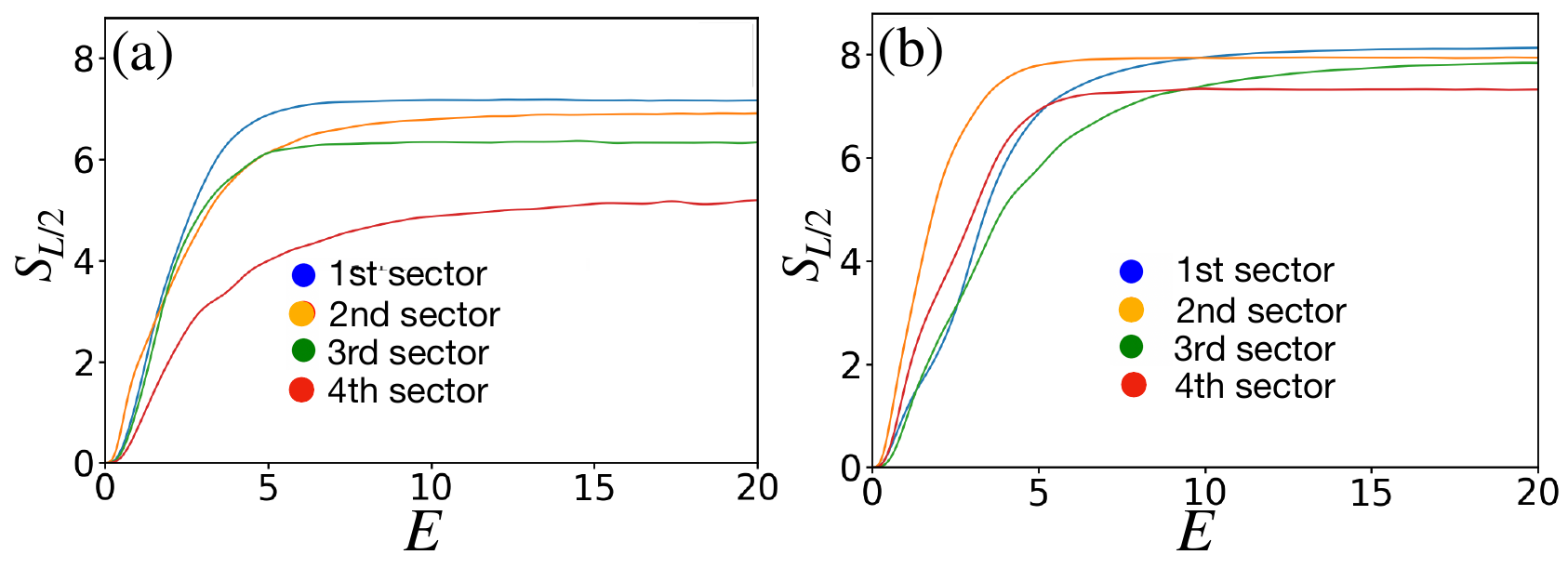}
  \caption{Growth of the entanglement entropy following a quench from an initial product state $|\psi_i\rangle$, chosen to belong to a specific Krylov subspace of the Hamiltonian, under (a) $H^{\mathcal{N}(1)}$ and (b) $H^{\mathcal{N}(2)}$. The long-time saturation value of the entanglement entropy increases with the dimension of the Krylov subspace accessed during the dynamics. Results are shown for $L=18$.}
  \label{fig:ent_quench}
\end{figure}

A qualitatively different picture emerges when we focus on \emph{dynamical entanglement generation}. Here, we prepare an initial product state $|\psi_i\rangle$ belonging to a specific Krylov subspace and study its unitary time evolution under $H^{\mathcal{N}(1)}$ and $H^{\mathcal{N}(2)}$. In contrast to the static eigenstate analysis, the long-time saturation value of the entanglement entropy is found to correlate strongly with the dimension of the Krylov subspace explored during the dynamics. As shown in Fig.~\ref{fig:ent_quench}, quenches within higher-dimensional Krylov subspaces lead to systematically larger entanglement entropy at long times, even though such a direct correspondence between Krylov space dimension and maximum entanglement is absent for $H^{\mathcal{N}(1)}$ at the level of individual eigenstates (revisit Fig.~\ref{fig:ent_h1_h2}(c)).

\begin{figure}
    \centering
    \includegraphics[width=0.5\textwidth]{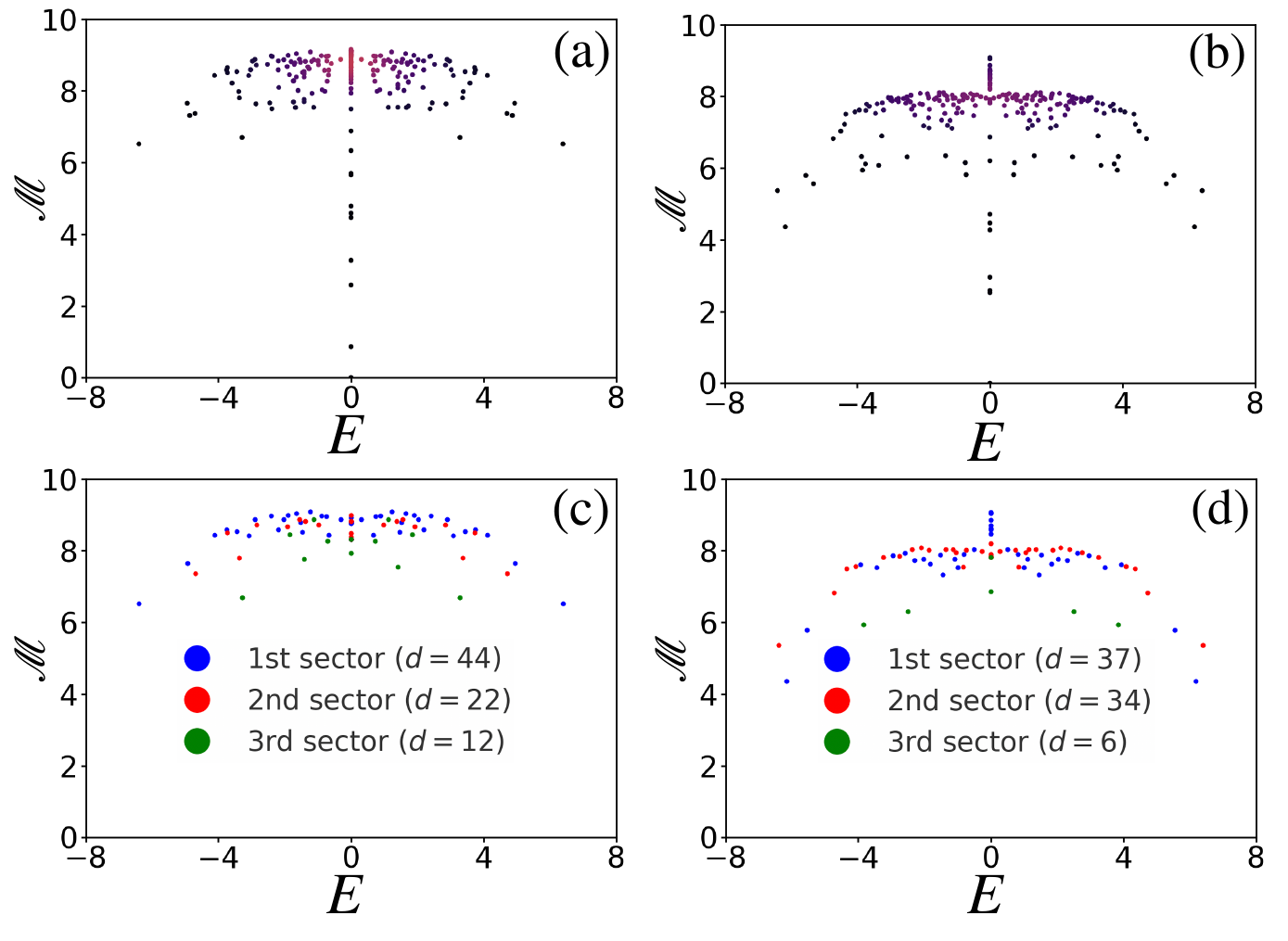} 
    \caption{Behavior of nonstabilzierness as quantified by Stabilizer R{\'e}nyi entropy ($\mathcal{M}$) for the eigenstates  belong to the symmetry resolved sectors of various kinetically constrained models we considered in our work, namely, (a) \(H^{\mathcal{N}(1)}\) (b) \(H^{\mathcal{N}(2)}\). Panels (c) and (d) show the SRE distribution for a few of the highest fragmented subspaces of   \(H^{\mathcal{N}(1)}\) and  \(H^{\mathcal{N}(2)}\), respectively. Here, all the plots are obtained for $L=12$.  } 
    \label{fig:SRE}
\end{figure}

\subsection{Stabilizer R{\'e}nyi Entropy}
The next aspect of quantum state preparation complexity discussed in our work is the nonstabilizerness. A useful way to quantify nonstabilizerness is by measuring how far a quantum state is from being a stabilizer state, which is referred to as the Stabilizer Rényi Entropy (SRE). For a pure state \(|\Psi\rangle\) of \(L\)  qubits, the SRE is defined as~\cite{Leone_2022}:

\begin{eqnarray}
\mathcal{M}(\Psi) = -\log_2 \left( \frac{1}{2^L} \sum_{P \in \mathcal{P}_L} |\langle \Psi | P | \Psi \rangle|^{4} \right),
\end{eqnarray} 
where \(\mathcal{P}_L\) is the set of all \(4^L\) Pauli strings constructed from the set \(\{I, \sigma^x, \sigma^y, \sigma^z\}\). The SRE is zero if and only if \(|\Psi\rangle\) is a stabilizer state. This measure remains unchanged under Clifford unitary operations and is additive when taking the tensor product of states. 
However, the computational cost of evaluating this expression scales as \(4^L\), which makes it extremely challenging to compute, even for moderately sized systems. 

In our study, we examine a maximum system size of \(L=12\) and plot the SRE behavior for all the considered models in Fig.~\ref{fig:SRE}. 
Similar to the behavior of entanglement as illustrated in Figs.~\ref{fig:ent_h1_h2} and \ref{fig:ent_qgl}, the SRE value obtained for the symmetry resolved fragmented sectors of \(H^{\mathcal{N}(1)}\) (Fig.~\ref{fig:SRE}(a)) and \(H^{\mathcal{N}(2)}\) (Fig.~\ref{fig:SRE}(b)) shows a more spread behavior than the Quantum Game of Life (Fig.~\ref{fig:ent_qgl_ppxpp}), which shows smooth behavior and remains nearly flat with variations in energy. This flatness indicates that the eigenstates in the middle of the spectrum exhibit some variation in terms of entanglement while sharing the same SRE content. This behavior resembles the scenario where quantum states remain maximally localized in the Fock basis, resulting in flat Shannon information entropy as discussed in~\cite{zhang_2025}. In case of \(H^{\mathcal{N}(1)}\) and \(H^{\mathcal{N}(2)}\), we separately plot the SRE behavior for a few of the largest subspaces to compare the SRE generating power of the dynamically disconnected Krylov sectors. Figure~\ref{fig:SRE}(c) reveals that for \(H^{\mathcal{N}(1)}\), the highest and the second-highest dimensional subspaces generate almost the same SRE value. This is different from the entanglement behavior we observed in Fig.~\ref{fig:ent_h1_h2}(c). Whereas, for \(H^{\mathcal{N}(2)}\), Fig.~\ref{fig:SRE}(d) shows many of the eigenstates of the second-highest subspace from the middle of the spectrum show higher SRE values than those obtained for the highest-dimensional sector.

\begin{figure}
  \includegraphics[width=.42\textwidth]{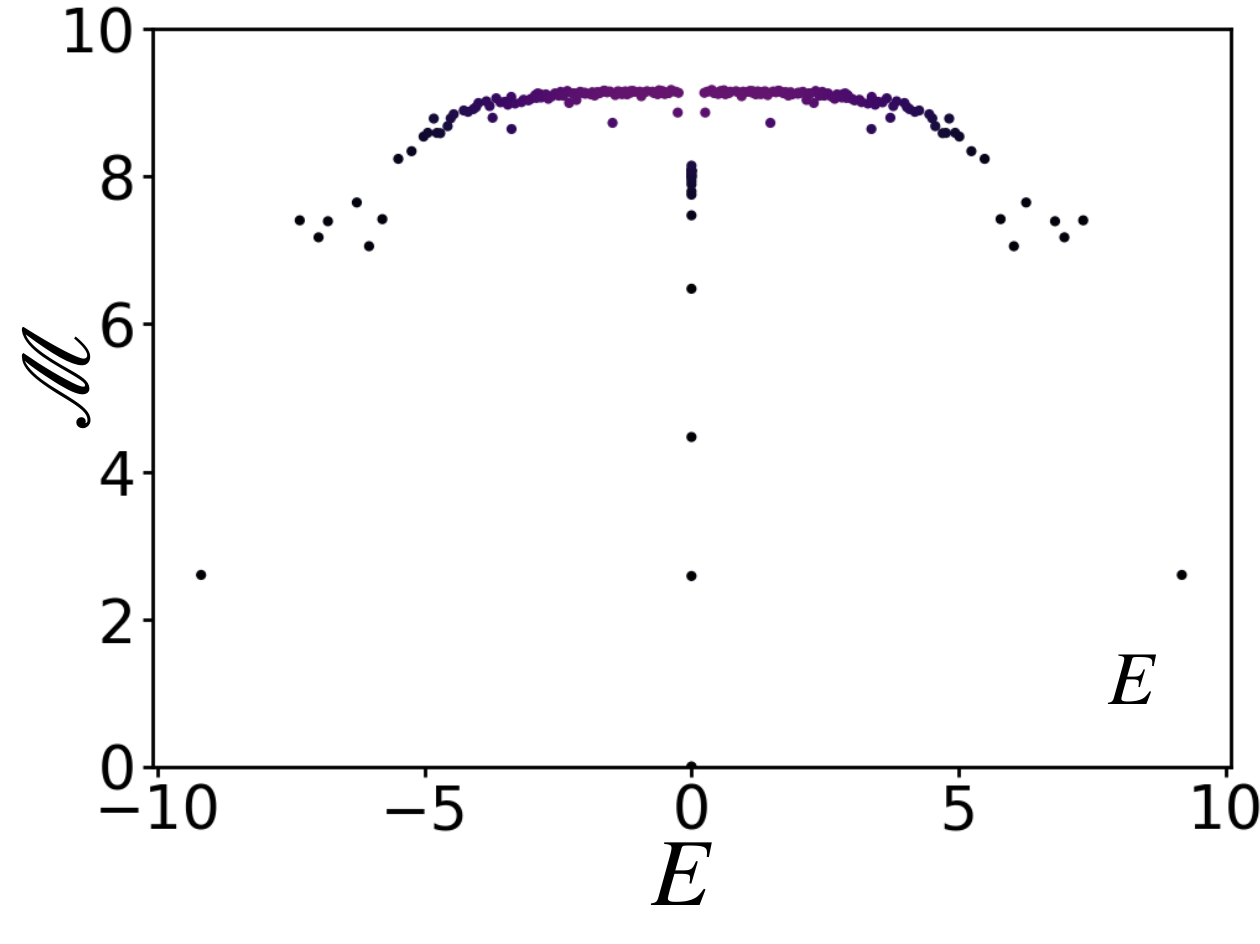}
  \caption{Behavior of the stabilizer R\'enyi entropy ($\mathcal{M}$) for eigenstates of the QGL model. We plot $\mathcal{M}$ for the eigenstates of $H^{\text{QGL}}$ in the zero-momentum, inversion-symmetric sector for $L=12$. For the bulk eigenstates the stabilizer R\'enyi entropy remains nearly independent of energy across the spectrum, indicating a uniform degree of non-stabilizerness among the eigenstates.}
  \label{fig:ent_qgl_ppxpp}
\end{figure}

However, in quenched dynamics, the saturation value of the SRE, similar to the entanglement entropy, is strongly correlated with the Krylov-subspace dimension for $H^{\mathcal{N}(1)}$, while for $H^{\mathcal{N}(2)}$ the saturation of the SRE becomes insensitive to the increase in the Krylov dimension, with the three largest subspaces producing nearly identical values as depicted in Fig.~\ref{fig:sre_quench}.

Our analysis therefore reveals that both the dynamical and static behaviors of these rule-based models exhibit a rich quantum-state–preparation complexity, as probed through entanglement and non-stabilizerness. In particular, for the HSF model, these complementary diagnostics can distinguish the underlying dynamics across different regimes, reflecting the distinct capacities of dynamically disconnected sectors to generate quantum resources.

\begin{figure}
  \includegraphics[width=.48\textwidth]{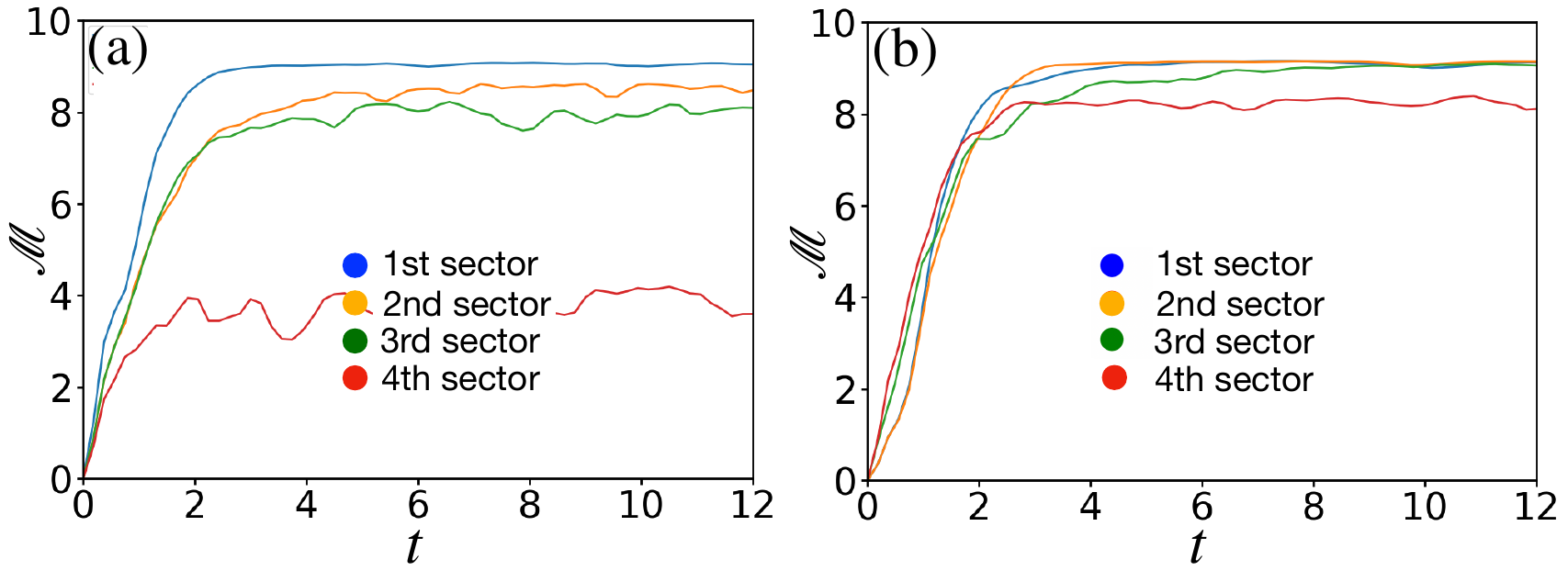}
  \caption{Growth of stabilizer R{\'e}nyi entropy following a quench from an initial product stabilizer state $|\psi_i\rangle$, chosen to belong to a specific Krylov subspace of the Hamiltonian, under (a) $H^{\mathcal{N}(1)}$ and (b) $H^{\mathcal{N}(2)}$. For $H^{\mathcal{N}(1)}$, the long-time saturation value of SRE increases with the dimension of the Krylov subspace accessed during the dynamics. However, for $H^{\mathcal{N}(2)}$, the first three highest Krylov spaces saturate to  almost same SRE value. Here we consider $L=12$.}
  \label{fig:sre_quench}
\end{figure}
\subsection{Chaotic behavior in the Model}

The behavior of entanglement entropy and SRE, though, unveils the complexity associated with the eigenstates of the model, it does not provide a direct insight into the complexity of the physical models that we have discussed above. Specifically, it does not directly inform us about whether these models are integrable or non-integrable. For that purpose, we consider probing the chaotic behavior associated with the model through a systematic approach that we describe in detail below. 

\subsubsection{Level Spacing Distribution}
\noindent 
We first study the energy level spacing distribution $P(s)$ where $s$ is the consecutive level spacing $s=e_{n+1}-e_n$ of the unfolded energy list. It is able to capture short range correlation in the spectrum. In our convention, $\{E_n\}$ is the original energy eigenvalue list and $\{e_n\}$ is the unfolded energy list. The unfolding is necessary to remove any kind of global effect in the spectrum due to the density of states of the system. This new energy set has an average level spacing  one. The unfolding process is explained in the Appendix~\ref{sec:ulfolding}.  Chaotic models exhibit behavior similar to that of random matrices. In these models, the phenomenon of level repulsion causes the probability density function \( P(s) \) to approach zero as the spacing \( s \) approaches zero. As a result, the distribution conforms to the Wigner-Dyson distribution. In contrast, integrable models display a different behavior due to the uncorrelated nature of their energy eigenvalues, leading to level spacing that follows a Poisson distribution. 

In our case, we compute the level spacing distribution for all the models discussed above and systematically unveil the chaotic behavior. As a first step, it is essential to resolve all types of symmetries and constraints  present in the model Hamiltonian. For example, in the case of \( H^{\text{QGL}} \), the zero-momentum inversion-symmetric sector (with \( \mathcal{I}=+1 \)) exhibits chaotic behavior, as shown in Fig.~\ref{fig_chaos_sff}(a). However, for \( H^{\mathcal{N}(1)} \) and \( H^{\mathcal{N}(2)} \), merely resolving the symmetries is insufficient; one must also examine the fragmented subspaces of the Hilbert space separately. 
For instance, with \( H^{\mathcal{N}(1)} \), in addition to focusing on the zero-momentum inversion symmetric space, we analyze the fragmented subspace with the largest dimension. The corresponding behavior is depicted in Fig.~\ref{fig_chaos_sff}(b).  Interestingly, for \( H^{\mathcal{N}(2)} \), the additional spin-flip symmetry has a significant impact on the level spacing behavior. If we do not take this symmetry into account and consider the fragmented subspaces collectively, the level spacing distributions show characteristics consistent with a Poisson distribution. This can be seen in Fig.~\ref{fig:H_2_unresolved}(a) in Appendix~\ref{sec:h2_extra}. 
However, including either the spin-flip symmetry or fragmentation alone is insufficient, as it does not yield the exact Wigner-Dyson shape of the level spacing distribution. The hidden Wigner-Dyson (WD) distribution can only be revealed by carefully resolving this symmetry and independently examining the fragmented subspaces. In our analysis, we focus on the \( \kappa= +1 \) sector, and the behavior of the level spacing distribution obtained from the largest Krylov subspace is presented in Fig.~\ref{fig_chaos_sff}(c).

\begin{figure}[htbp]
    \centering
    \includegraphics[width=0.48\textwidth]{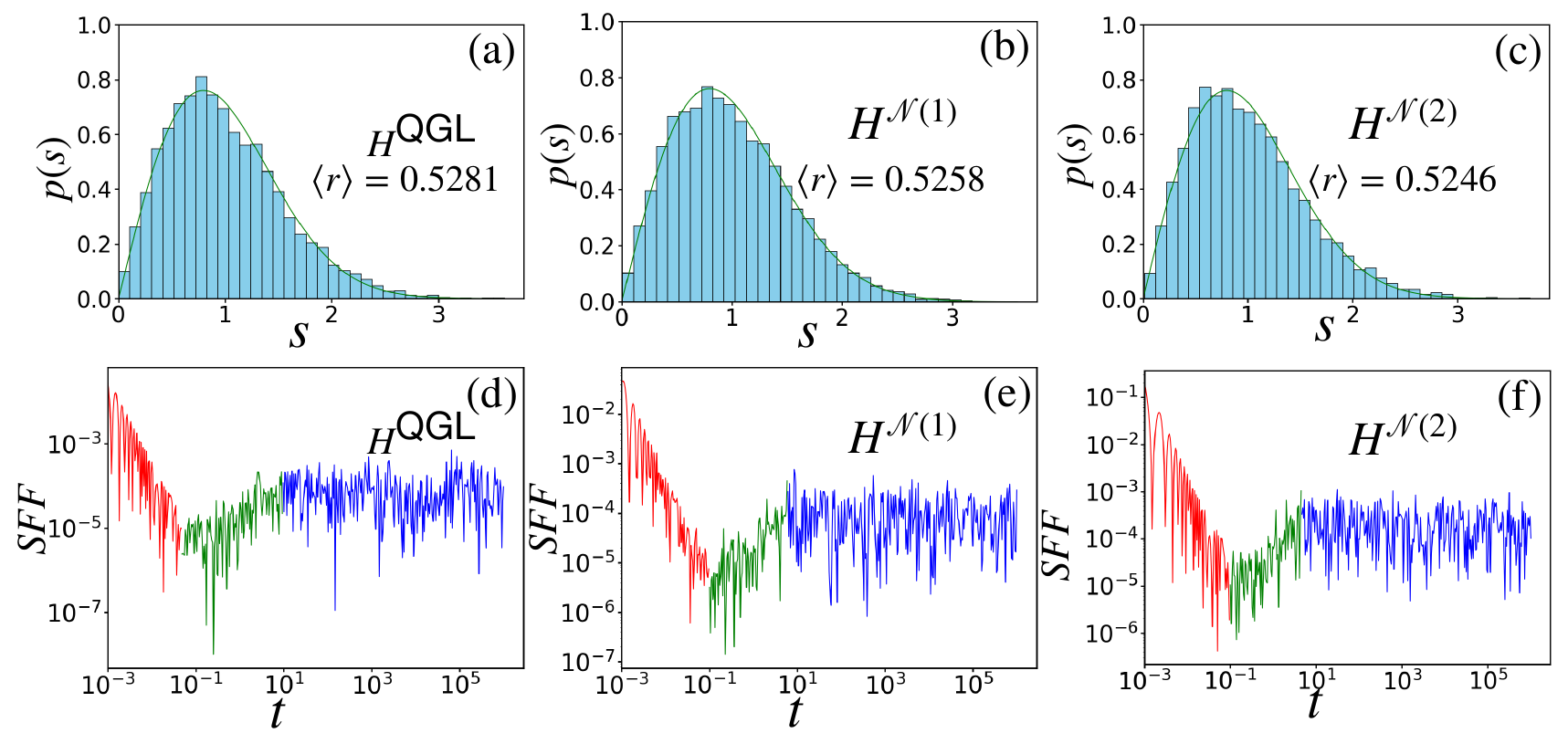} 
    \caption{The level spacing distributions along with the values of level spacing ratios $\langle r \rangle$, for (a) \( H^{\text{QGL}} \) (with \( L = 20 \)), (b) \( H^{\mathcal{N}(1)} \) (with \( L = 24 \)) and (c)  \( H^{\mathcal{N}(2)} \) (with \( L = 22 \)) are analyzed.  If the total number of eigenvalues in a sector is \( D_{e} \), the energy list \( \{e_i\} \) is used, where \( i \) ranges from \( [D_{e}/10] \) to \( [D_{e}/2 - 500] \). For both \( H^{\text{QGL}} \) and \( H^{\mathcal{N}(1)} \), the zero momentum inversion symmetric sector is considered (for \( H^{\mathcal{N}(1)} \) we further consider the largest connected subspace). In contrast, for \( H^{\mathcal{N}(2)} \), the analysis focuses on the largest connected sector comprised of both zero momentum inversion symmetry as well as the spin flip symmetry.  In the bottom figures, we illustrate the behavior of the Spectral Form Factor (SFF) for the same models and system sizes, (d) \( H^{\text{QGL}} \) (with \( L = 20 \)), (e) \( H^{\mathcal{N}(1)} \) (with \( L = 24 \)) and (f) \( H^{\mathcal{N}(2)} \) (with \( L = 22 \)).  The structure characterized by a slope, dip, ramp, and plateau is clearly visible in each plot. We mark the slope, ramp, and plateau using red, green, and blue colors, respectively. Note that, except in the computation of the level spacing ratio, the unfolded spectra of the various symmetry-resolved sectors are considered in all cases.}
    \label{fig_chaos_sff}
\end{figure}


\subsubsection{Level spacing ratio}\noindent
To obtain a more accurate characterization of the level spacing behavior, we additionally compute the level spacing ratio. It is represented by the average value \(\langle r \rangle\), where \(r\) is defined as follows:
\begin{eqnarray}
r = \text{min} \left\{r_n, \frac{1}{r_n} \right\}, \quad r_n = \frac{E_{n+1} - E_n}{E_n - E_{n-1}}.
\end{eqnarray}
The average is calculated over all eigenvalues. In this analysis, we use the original energies \(E_n\) rather than the unfolded energies \(e_n\), because any global effects from the system's density of states will cancel out in the ratio. 

According to random matrix theory, integrable models have an average value of \(\langle r \rangle \approx 0.3863\), while chaotic models yield \(\langle r \rangle \approx 0.5307\). In our study, for the models \(H^{\text{QGL}} \ (L=20)\), \(H^{\mathcal{N}(1)} \ (L=24)\), and \(H^{\mathcal{N}(2)} \ (L=22)\), the computed level spacing ratios are 0.5281, 0.5258, and 0.5246, respectively. 

\subsubsection{Spectral Form Factor}
To gain a finer look at the spectral properties of a model, we further aim to study the long-range spectral correlations, which cannot be captured by level spacing alone. In this case, the Spectral Form Factor (SFF)~\cite{Hikami_1997} is one of the simplest and most useful quantities to calculate. It helps us to understand the long-range, universal fluctuations in the energy spectrum of quantum systems. Because of this, the SFF has become important in many areas of physics — from the semi-classical study of quantum chaos~\cite{Bertini_2018}, to black hole physics~\cite{Choi_2023}, many-body quantum chaos~\cite{Dong_2025}, and related phenomena.
The Spectral Form Factor (SFF) is defined as 
\begin{equation}
    \text{SFF} = \frac{1}{N^2} \sum_{m,n=1}^N e^{i(e_m-e_n)t},
\end{equation}
where $N$ is the total number of energy eigenvalues. Here, we have considered the unfolded energies. In chaotic models, the evolution of the SFF typically shows three distinct features: an early-time slope, an intermediate dip–ramp region, and a long-time plateau. Among these features, the ramp reflects the presence of long-range energy correlations in the model.
Fig.~\ref{fig_chaos_sff} depicts the behavior for  (d) $H^{\mathrm{QGL}}$,  (e) $H^{\mathcal{N}(1)}$, (f) $H^{\mathcal{N}(2)}$. As mentioned earlier, three distinct features that serve as markers of chaotic behavior are clearly visible in all cases. For clarity, these regions are highlighted using three different colors.

Our analysis also shows the advantage of using the Spectral Form Factor (SFF) compared to level statistics. As noted earlier, the additional spin-flip symmetry in the model $H^{\mathcal{N}(2)}$ requires an extra step of symmetry resolution. Without this step, the level spacing distribution deviates strongly from the Wigner--Dyson form and instead resembles a Poisson distribution, as seen in Fig.~\ref{fig:H_2_unresolved}(a) in Appendix~\ref{sec:h2_extra}. In contrast, the SFF still displays the characteristic ramp behavior even without resolving this last symmetry. See Fig.~\ref{fig:H_2_unresolved}(b). This suggests that, for quantum many-body models where symmetry resolution plays a crucial role in identifying genuine chaotic behavior, the SFF can provide a more reliable diagnostic.

\begin{figure}[htbp]
    \centering
    \includegraphics[width=0.48\textwidth]{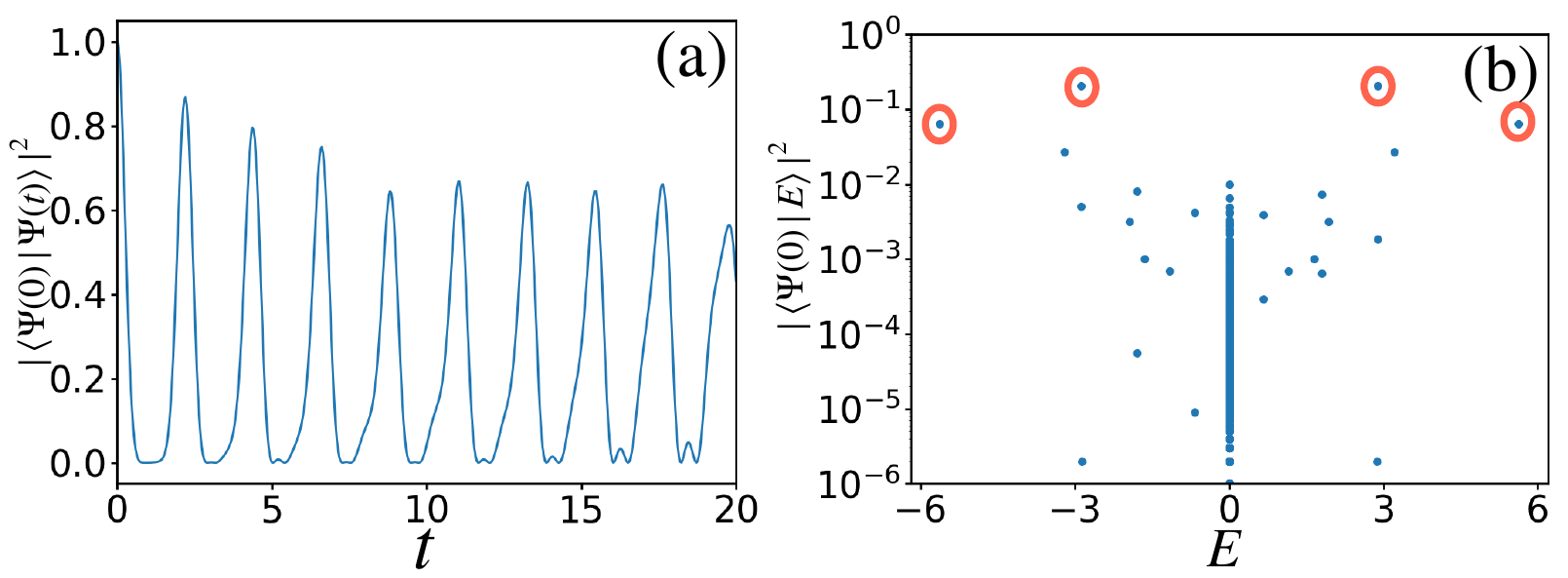} 
    \caption{Behavior of return probability and overlap with energy eigenstates of an initial state  $|K=6\rangle^{\otimes 2}$ as defined in Eq.~(\ref{eqn:scar_state}) when quenched using Hamiltonian $H^{\mathcal{N}(0)}$. (a) shows a clear revival of the state signaturing existence of quantum scars that can also be validated further by taking the overlap of $|K=6\rangle^{\otimes 2}$ with all the eigenstates of the model as shown in (b). Here, we can clearly see that four eigenstates exist, sharing a relatively higher value of the overlap (marked using red circles), which characterizes the typical scar states. This behavior is exactly the same as obtained in Fig.~{\color{red}1}(a) of Ref.~\cite{Kerschbaumer_2025} for the conventional PPXPP model with Rydberg constraint as the largest Krylov space  in both the cases coincides. Both the plots here are obtained for $L=12$.} 
    \label{fig_scar_h0}
\end{figure}

\begin{figure}[htbp]
    \centering
    \includegraphics[width=0.5\textwidth]{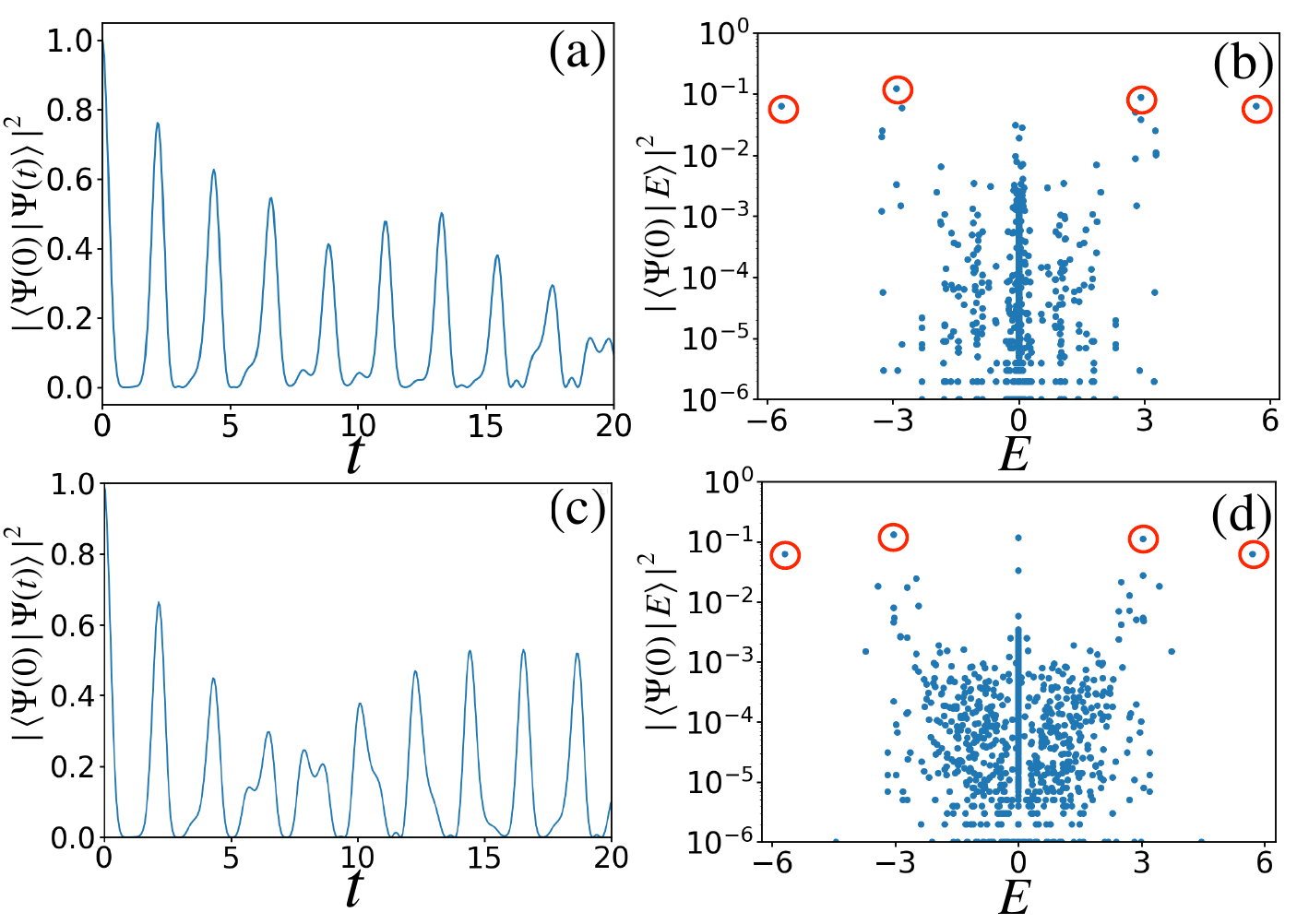} 
    \caption{Illustration of the existence of quantum scars in fragmented subspaces of the model \( H^{\text{Pert}}_{\text{PPXPP}}(k) \). Similar to the marker of quantum scars depicted in Fig.~\ref{fig_scar_h0}, we notice that  (a) and (b) show similar behavior for $H^{\text{Pert}}_{\text{PPXPP}}(k=1)$ with $\delta = 0.09$, and (c)-(d) depicts the same for $H^{\text{Pert}}_{\text{PPXPP}}(k=2)$ with $\delta = 0.50$. Here we considered $L=12$.}
    \label{fig_scar_Augmented}
\end{figure}

\section{Quantum scars in fragmented subspaces}
\label{sec:quantum scars}

 Appearance of kinetically constrained models that consist of a set of atypical states that do not thermalize is not a new phenomena. In literature, the existence of such models have been well studied in PXP and related family of models~\cite{Serbyn_2021,Turner_2018_NatPhys,Turner_2018_PRB,Bernien_2017,Mohapatra_2023, Jamir_2024} and recently, some prescriptions have also been suggested to find whether quantum many-body scar exists~\cite{Han_2023,ren2025}. However, for a given many-body model, finding scar is a nontrivial task. 
 
 In our work, the entanglement behavior observed for $H^{\mathcal{N}(1)}$, $H^{\mathcal{N}(2)}$, $H^{\text{QGL}}$ and  does not provide evidence for the presence of quantum many-body scars. Additional support for this conclusion comes from the absence of revival dynamics when standard initial states such as $|\mathbb{Z}_2\rangle$ and $|\mathbb{Z}_4\rangle$, commonly used in previous studies, are employed. However, we demonstrate that when these models are combined with $H^{\mathcal{N}(0)}$, signatures of persistent nonthermal dynamics survive even in the strong perturbation limit. Before presenting these results, we remind the reader that the largest Krylov subspace of $H^{\mathcal{N}(0)}$ coincides with the conventional PPXPP model  with Rydberg constraints for which the characteristic features of quantum scar states are already known. For completeness, we first reproduce this behavior by examining the time evolution of the overlap $\langle \Psi(t)|\Psi(0)\rangle$ for an appropriate initial state evolved under $H^{\mathcal{N}(0)}$, as reported in Fig.~1(a) of Ref.~\cite{Kerschbaumer_2025}. Here, $|\Psi(0)\rangle=|K=6\rangle^{\otimes 2}$  with
 \begin{eqnarray}
 |K\rangle = |0\rangle^{\otimes K/2} \otimes |W\rangle_{K/2},
 \label{eqn:scar_state}
 \end{eqnarray}
where the first half of the unit cell is prepared in the product state
$ 
|0\rangle^{\otimes K/2} = |0 0 \cdots 0 \rangle, 
$ 
and the second half hosts the entangled $W$-state,  given by 
{\small
\begin{eqnarray}
|W\rangle_{K/2} = \sqrt{\frac{2}{K}} \Big( |1 0 \cdots 0\rangle 
+ |0 1 \cdots0\rangle 
+ \cdots 
+ |0\cdots 0 1 \rangle \Big),\nonumber
\end{eqnarray}
} 
\noindent which corresponds to a uniform superposition with a single excitation $|1\rangle$ across the $K/2$ sites of the second half of the unit cell. Figure \ref{fig_scar_h0} shows strong revivals in the behavior of $\langle \Psi(t)|\Psi(0)\rangle$, signifying the existence of quantum scar states that we also confirm by taking the overlap of $|K=6\rangle^{\otimes 2}$ with the eigenstates of the model.\\

\begin{figure}[htbp]
    \centering
    \includegraphics[width=0.5\textwidth]{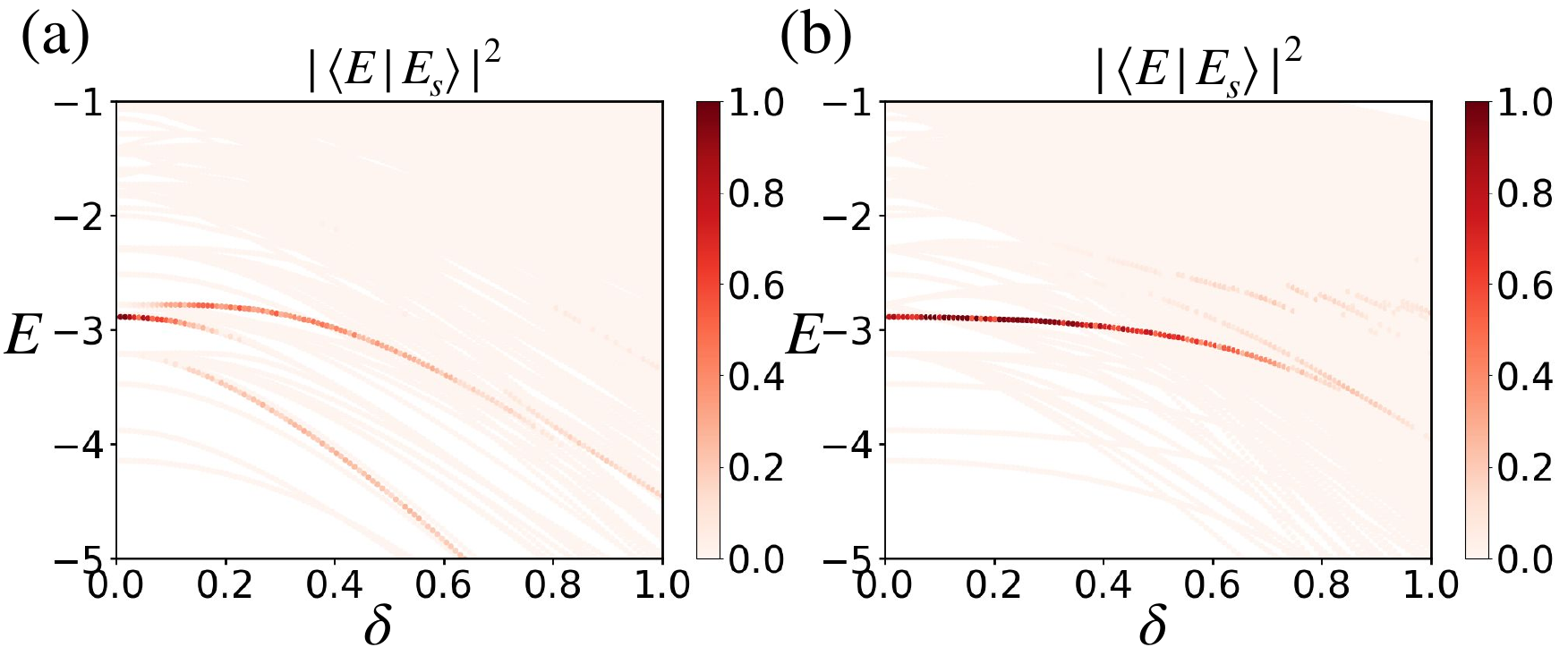} 
    \caption{Intensity plot of the squared overlap between the exact eigenstates $|E\rangle$ of the perturbed Hamiltonian
\( H^{\text{Pert}}_{\text{PPXPP}}(k) \) and the exact scar state $|E_s\rangle$ of \( H^{\mathcal{N}(0)} \) for (a) \( k = 1 \)
and (b) \( k = 2 \), respectively. The results are consistent with those shown in
Fig.~\ref{fig_scar_Augmented}. In both cases, beyond a critical value of the detuning
parameter \( \delta \) (e.g., $\delta\lesssim0.09$ for $k=1$ and $\delta\lesssim 0.50$ for $k=2$), the scar disappears, as the probe state develops significant
overlap with other eigenstates (away from the bulk) as well.}
    \label{fig:scar_overlap}
\end{figure} 
We next study the robustness of the scar behavior under perturbations. In particular, we perturb the model by adding $H^{\mathcal{N}(k)}$ and consider the Hamiltonian
\begin{equation}
H^{\text{Pert}}_{\text{PPXPP}}(k) = H^{\mathcal{N}(0)} + \delta \cdot H^{\mathcal{N}(k)} .
\end{equation}
We find that the scar signatures persist up to finite perturbation strengths, namely $\delta \approx 0.09$ for $k=1$ and $\delta \approx 0.50$ for $k=2$, as shown in Fig.~\ref{fig_scar_Augmented}. The robustness of the scar is illustrated more explicitly in Fig.~\ref{fig:scar_overlap}, where in both cases in the respective $\delta$ range (e.g., $\delta\lesssim0.09$ for $k=1$ and $\delta\lesssim 0.50$ for $k=2$), a single eigenstate of the perturbed Hamiltonian  can be continuously traced back to the scar state of the unperturbed model. This can be realized when we examine the squared overlap between the
exact eigenstates of the perturbed Hamiltonian  $H^{\text{Pert}}_{\text{PPXPP}}(k)$ ($|E\rangle$)
and the exact scar state of $H^{\mathcal{N}(0)}$ ($|E_s\rangle$).

\begin{table}[t]
\centering
\small
\begin{tabular}{p{2.50cm} p{2.50cm} p{2.50cm}}
\hline\hline
{\bf Hamiltonian} & {\bf Fragmentation} & {\bf Scarring} \\
\hline
$H^{\mathcal{N}(0)}$
& Trivial HSF
& Strong scars \\
\hline
$H^{\mathcal{N}(1)}$
& Strongest  HSF
& No scars \\
\hline
$H^{\mathcal{N}(0)} + \delta \cdot H^{\mathcal{N}(1)}$
& Strong HSF
& Weak scars \\[-2pt]
& 
& ($\delta \lesssim 0.09$) \\
\hline
$H^{\mathcal{N}(2)}$
& Weak   HSF
& No scars \\
\hline
$H^{\mathcal{N}(0)} + \delta \cdot H^{\mathcal{N}(2)}$
& Weak  HSF
& Strong scars \\[-2pt]
& 
& ($\delta \lesssim 0.50$) \\
\hline
$H^{\text{QGL}}$ 
& No    HSF
& No scars \\
\hline
$H^{\text{Tot}}$ 
& No    HSF
& No scars \\
\hline
\end{tabular}
\caption{Summary of Hilbert-space fragmentation (HSF) and quantum many-body scarring behavior for family of KCMs we have explored in our work. The table highlights the coexistence of fragmentation and scarring, as well as their distinct stability regimes under controlled perturbations. Here, \emph{weak scars} denote scar signatures that vanish under arbitrarily small perturbations $\delta\gtrsim 0.09$ as can be seen for  $H^{\mathcal{N}(1)}$, whereas for $H^{\mathcal{N}(2)}$ the scarring behavior remains robust and survives up to perturbation strengths of approximately $\delta \lesssim 0.5$. Similarly, as mentioned in the main text, here \emph{trivial} HSF refers to the case where, although the condition $d_L/\mathcal{D}_L \rightarrow 0$ as $L \gg 1$ is formally satisfied, most of the Krylov subspaces consist of a single basis state that is trivially annihilated by the Hamiltonian due to the kinetic constraint.}
\label{Table:H_ppxpp_pert}
\end{table}

These observations demonstrate that $H^{\text{Pert}}_{\text{PPXPP}}$ provides an explicit example of a system that simultaneously hosts Hilbert-space fragmentation and quantum many-body scars. Notably, as we observed from Fig.~\ref{fig:scar_overlap}, the scarred eigenstates can be continuously traced back to the unperturbed limit, for some values of $\delta$, implying that, in terms of quantum scars, the perturbations do not qualitatively change the physics from the unperturbed case. 

  All these observations made above combining the behavior of Hilbert space  fragmentation and quantum many-body scar help us in categorizing these models in three cases: (i) $H^{\mathcal{N}(0)}$ exhibits strong scarring without nontrivial fragmentation (the Krylov subspaces are dominated by sectors consisting of a single basis state, which is annihilated by the Hamiltonian due to the kinetic constraint). (ii) $H^{\text{Pert}}_{\text{PPXPP}}(k=1)$ shows strong fragmentation with weak but stable scarring, which survives up to $\delta \lesssim 0.09$. (iii) $H^{\text{Pert}}_{\text{PPXPP}}(k=2)$ displays relatively weaker  fragmentation than $H^{\text{Pert}}_{\text{PPXPP}}(k=1)$  together with comparatively more robust scarring, persisting up to $\delta \lesssim 0.50$. This establishes a regime in which three distinct features: scarring, fragmentation, and perturbative stability coexist within a single controlled framework. We summarize these results in in Table~\ref{Table:H_ppxpp_pert}.

\section{Quantum complexity study of ${H^{\mathrm{Tot}}}$}
\label{sec:Htot}
\begin{figure}
    \centering
    \includegraphics[width=0.45\textwidth]{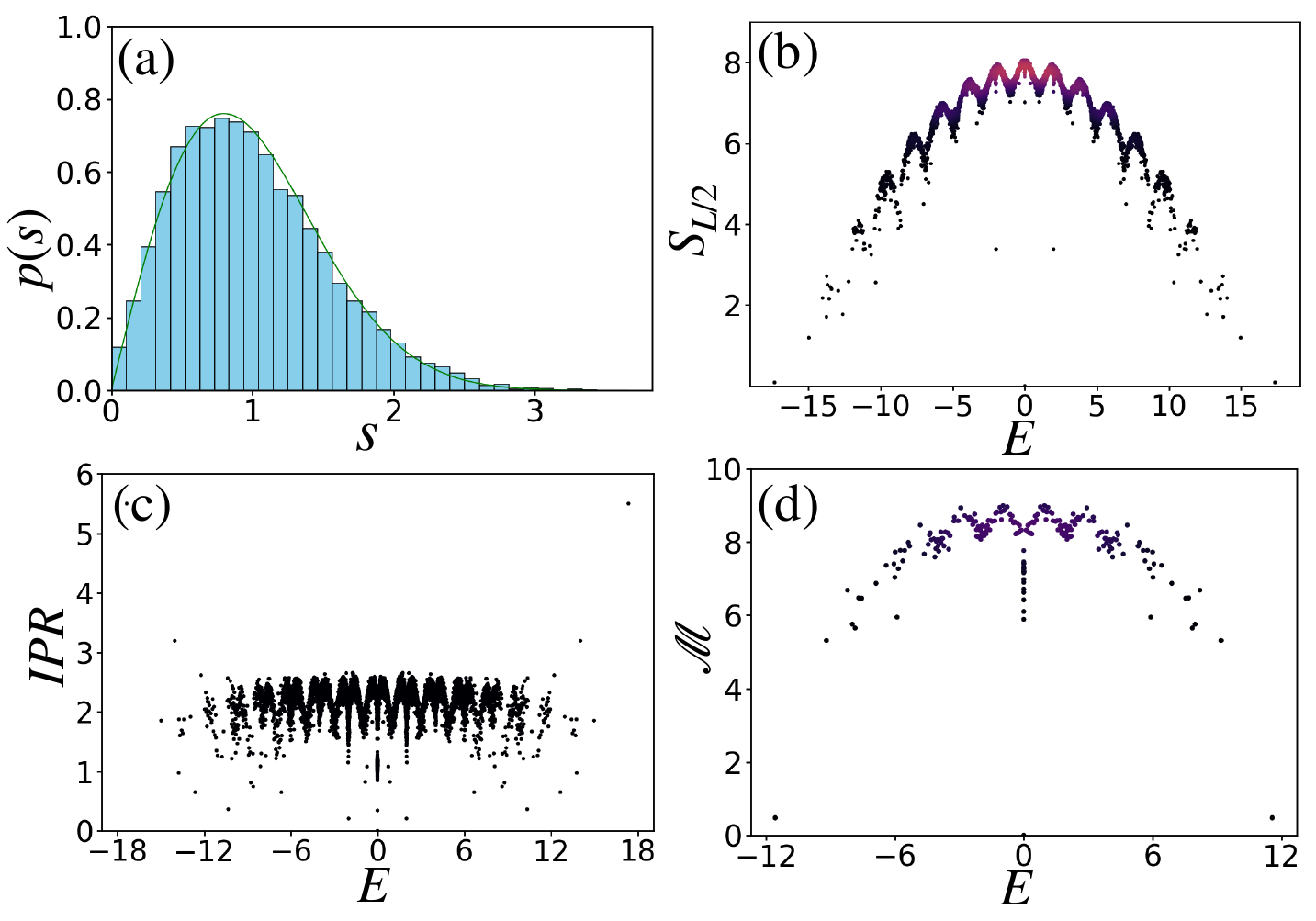} 
    \caption{Quantum complexity for the Hamiltonian \(H^{\mathrm{Tot}}\), as defined in Eq.~(\ref{eqn:main_Ham}). (a) Ilustrates the chaotic behavior of the  model for \(L=20\).(b) Shows the behavior of entanglement obtained for $L=18$  remains notably distinct from that obtained for other KCMs discussed previously. Notably, in this case, in addition to the primary arch-like behavior, the presence of low-entangled states creates a sub-arch structure or an entanglement band.
Similar patterns are observed in the plots shown in (c) and (d), where we plot the behavior of the Inverse Participation Ratio (IPR) for \(L=18\) and the Stabilizer R{\'e}nyi Entropy,  \(\mathcal{M}\) for \(L=12\), respectively. In case of (b), the y-axis is rescaled by a factor $10^{-3}$.}
    \label{fig_H_tot}
\end{figure}
We devote this section to discuss the quantum complexity properties of the Hamiltonian \(H^{\mathrm{Tot}}\), separately. Interestingly, in this case, we observe that although the model is chaotic like all the KCMs discussed previously (see Fig.~\ref{fig_H_tot}(a)) and does not exhibit any  fragmentation, its quantum complexity behavior shows a distinct feature from all the other models we have discussed so far. The first distinct feature we highlight here is the EE behavior with energy as shown in Fig.~\ref{fig_H_tot}(b), displaying sub-arch-like splitting of the primary arch shape, resulting in an entanglement band-like structure. This behavior we find counterintuitive, and it does not translate immediately from all other observed behavior. To substantiate our claim, in Fig.~\ref{fig_H_tot}(c), we also plot the Inverse Participation Ratio (IPR)\footnote{Note also that in the literature,  the quantity $\sum_i |\langle i | \Psi\rangle|^4$ is also often referred to as the inverse participation ratio (IPR).}, defined as 
\[
\mathrm{IPR} = \frac{1}{\sum_i |\langle i|\Psi\rangle|^4},
\]
with \(\{|i\rangle\}\) being the basis states in the symmetry reduced sector. The figure shows that IPR exhibits the same behavior as EE.  IPR is a standard diagnostic of localization in Hilbert space where larger values indicate stronger delocalization~\cite{Evers_2008}. In our case,  it exhibits the same band-like structure as the EE implies that the eigenstates are not uniformly delocalized, despite the system being chaotic. Instead, different regions of the spectrum host states with systematically varying degrees of delocalization. As a result, the observed entanglement band structure is not a random fluctuation but reflects an underlying organization of eigenstates in Hilbert space. The same imprint can be found in SRE as depicted in Fig.~\ref{fig_H_tot}(d).  Hence, although the system does not exhibit conventional fragmentation, it indicates that EE and other measures are not smooth functions of energy. Even in the bulk, there exist eigenstates that cause dips in the entanglement and SRE and create this band-like structure.

This observation motivates a closer investigation of the origin of these entanglement ripples. Notably, the total Hamiltonian can be written as
\begin{equation}
H^{\mathrm{Tot}} = \sum_i \sigma^x_i - H^{\mathcal{N}(4)}.
\end{equation}
To systematically explore how these ripples emerge, we introduce a tunable Hamiltonian
\begin{equation}
H^{\mathrm{Tot}}(\Delta) = \sum_i \sigma^x_i - \Delta \cdot  H^{\mathcal{N}(4)}.
\end{equation}
Figure~\ref{fig_H_tot_Delta} shows the behavior of the entanglement entropy for different values of $\Delta$. For $\Delta = 0$, the system reduces to the undressed model ``$H_x=\sum_i \sigma^x_i$", which is integrable and exhibits localized entanglement behavior with well-separated bands. As soon as $\Delta$ is turned on, these localized bands begin to hybridize, leading to partial mixing between neighboring eigenstates. With increasing $\Delta$, the entanglement develops a ripple-like structure that we observe in the full $H^{\mathrm{Tot}}$ model at ($\Delta = 1$) as also depcited in Fig.~\ref{fig_H_tot}(a).

Interestingly, for sufficiently large values of $\Delta$, the ripples are completely washed out and the entanglement entropy becomes a smooth function of energy, consistent with ergodic behavior. This crossover highlights how weak integrability-breaking perturbations can generate nontrivial structure in eigenstate entanglement, even in the absence of strict fragmentation.

\begin{figure}
\centering
\includegraphics[width=0.5\textwidth]{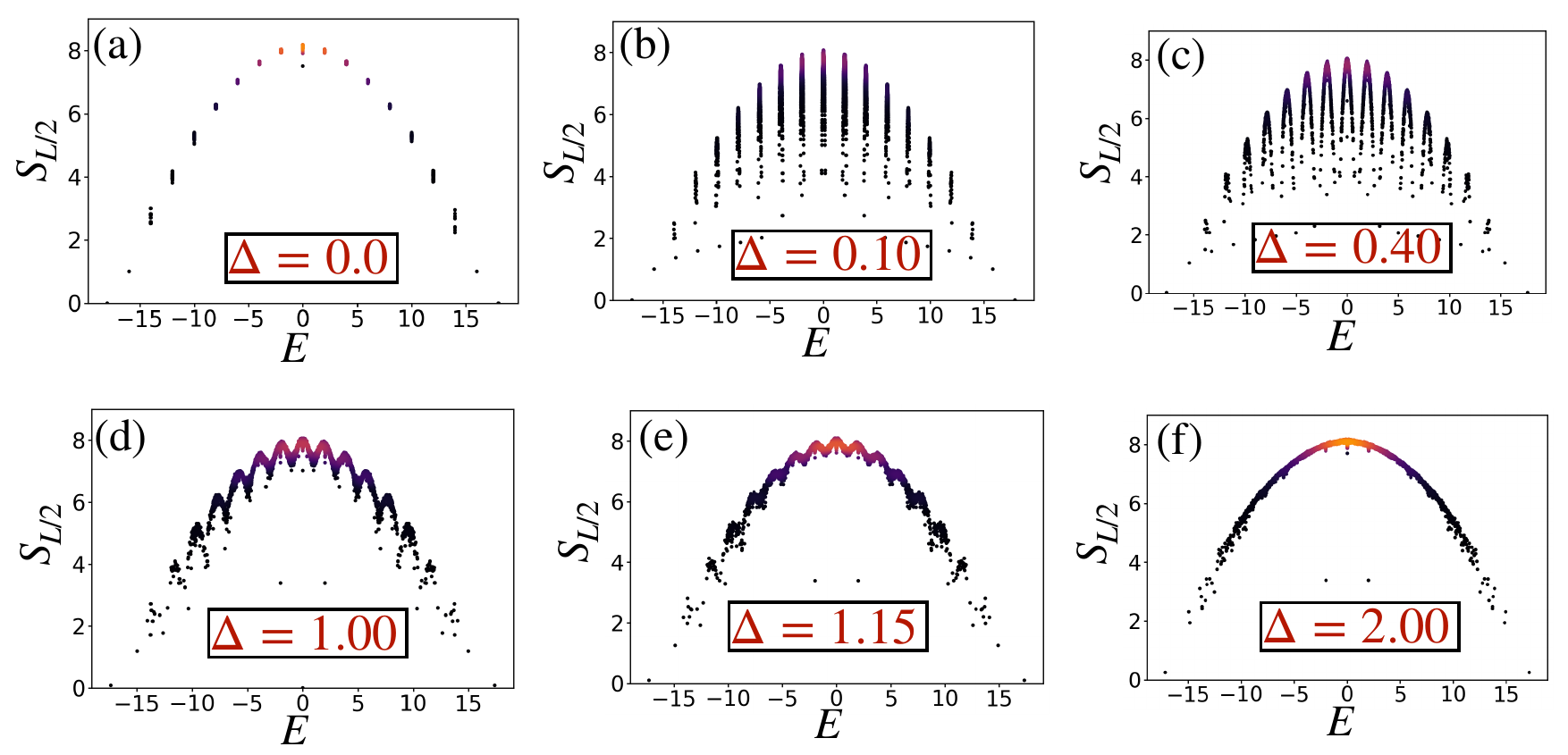}
\caption{The entanglement entropy behavior of $H^{\mathrm{Tot}}(\Delta)$. This plot supports the behavior of EE obtained in Fig.~\ref{fig_H_tot}. Here we analyze the entanglement entropy as a function of energy for the tunable Hamiltonian ($H^{\mathrm{Tot}}(\Delta) = \sum_i \sigma^x_i - \Delta H^{\mathcal{N}(4)})$. Starting from $\Delta = 0$, corresponding to the undressed and integrable $\sigma_i^x$ model, the spectrum exhibits sharply defined entanglement bands. As $\Delta$ is increased, these bands begin to hybridize, leading to the emergence of ripple-like structures in the entanglement profile. For sufficiently large $\Delta$, the ripples are completely washed out and the entanglement entropy becomes a smooth function of energy.}
\label{fig_H_tot_Delta}
\end{figure}

\section{Discussion}
In this work, we have investigated quantum complexity in a family of rule-based kinetically constrained models, examining both the challenges of state preparation and the growth of complexity under dynamical evolution. The family  includes  the celebrated quantum Game of Life and their variations.  A key characteristic of this class of models is that it goes beyond the strict blockade condition encountered in the PXP family, which is realized through symmetric projectors. Instead, the quantum Hamiltonians considered here involve terms constructed from asymmetric projectors as well. We have shown how entanglement, nonstabilizerness, and  quantum chaos  interplay with each other in these models. In particular, we  have unveiled the chaotic behavior through systematic symmetry resolutions of  the models. However, there are cases where mere symmetry resolution has not  revealed the chaotic behavior, as the Hilbert space exhibits dynamically  disconnected sectors, which has resulted in Hilbert space fragmentation. Along with that, for some of the models, the fragmented subspaces have further hosted  quantum many-body scars. This has given us a scope to extract non-trivial physics  that exhibits quantum chaos, quantum many-body scars, as well as Hilbert space  fragmentation out of a single family of quantum many-body models. 

We have further characterized the fragmented subspaces by their resource  generation capabilities. In particular, we have studied how the dimension of the  Krylov subspaces correlates with the maximum entanglement entropy generation as  well as their classical simulability aspects as quantified by Stabilizer R{\'e}nyi entropy. Our analysis reveals that the static and dynamical behaviors of these models exhibit distinct capacities for quantum resource generation, which can be instrumental in their characterization. Hence, our study sheds light on the interplay between dynamical constraints, accessible Krylov subspaces, and the resulting complexity of quantum state preparation and evolution for structurally similar  kinetically constrained models that can be instrumental for further fundamental  as well as application-based studies. For example, in our work on QGL and related models, we have not found evidence of quantum many-body scars, which we  plan to pursue more systematically in future work. Along with that, whether the  quantum many-body scar or the fragmented phase can be applied for practical  applications such as quantum sensing is another direction we also wish to explore. Another aim will be to characterize the fragment subspaces in terms of their Krylov Complexity behavior~\cite{Rabinovici_2022,Budhaditya_2022,Budhaditya_2025,Balasubramanian2025}.  Finally, the entanglement localization feature imprinted in all the quantum properties we obtained for the model $H^{\mathrm{Tot}}$ is another aspect that we will study in detail in our future work. 
\label{Sec discussion}

\acknowledgments{S.S.R. wants to thank Germ{\'a}n Sierra for introducing the QGL model and sharing valuable comments on the manuscript. We also want to thank Philipp Hauke, Himadri Shekhar  Dhar, Anindita Bera  for reading the manuscript and sharing useful comments. We thank Soumik Bandyopadhyay for useful discussion on spectral form factor. S.S.R. acknowledges financial support from the Faculty Research Scheme, IIT (ISM) Dhanbad, India, under Project No. FRS/2024/PHYSICS/MISC0110, and from the Anusandhan National Research Foundation (ANRF), Government of India, under Grant Nos. ANRF/ARG/2025/004617/PS and ANRF/ECRG/2025/002793/PMS. We also acknowledge the use of the QuSpin package~\cite{QuSpin} for the numerical simulations presented  in this work. }

\appendix

\begin{widetext}

\section{Effect of open boundary conditions}
\label{sec:obc}
The presence of open boundary conditions imposes additional constraints on the evolution of the basis states, thereby enhancing the fragmentation of the Hilbert space. To illustrate this, we focus on two models where, in the case of periodic boundary condition (PBC), no fragmentation was reported, namely, the quantum game of life $(H^{\text{QGL}})$ and the $H^{\text{Tot}}$ models. However, when open boundary conditions are imposed (along with inversion symmetry), we find that both models exhibit {\bf exactly 10 fragmented sectors}. These sectors are characterized by specific configurations of the end spins, labelled by  ($s_1,s_2, s_{N-1},s_N$), given by
\begin{align}
& {\bf 1.}\, |0_1 0_2 \cdots 0_{N-1} 0_N\rangle, 
{\bf 2.}\, |1_1 1_2 \cdots 1_{N-1} 1_N\rangle,  {\bf 3.}\, |1_1 0_2 \cdots 0_{N-1} 1_N\rangle,
{\bf 4.}\, |0_1 1_2 \cdots  1_{N-1} 0_N\rangle, \\
& {\bf 5.}\, |0_1 1_2 \cdots 0_{N-1} 1_N\rangle + |1_1 0_2 \cdots  1_{N-1} 0_N\rangle , {\bf 6.}\, |1_1 1_2 \cdots 0_{N-1} 0_N\rangle + |0_1 0_2 \cdots 1_{N-1} 1_N\rangle,\nonumber \\
& {\bf 7.}\, |0_1 1_2 \cdots 1_{N-1} 1_N\rangle  +  |1_1 1_2 \cdots 1_{N-1} 0_N\rangle , {\bf 8.}\, |1_1 0_2 \cdots 1_{N-1} 1_N\rangle +  |1_1 1_2 \cdots 0_{N-1} 1_N\rangle ,\nonumber \\
& {\bf 9.}\, |0_1 0_2  \cdots 0_{N-1} 1_N\rangle  + |1_1 0_2  \cdots  0_{N-1} 0_N\rangle, {\bf 10.}\,
|0_1 0_2  \cdots 1_{N-1} 0_N\rangle  + |0_1 1_2\cdots 0_{N-1} 0_N\rangle .\nonumber
\end{align}
 For the QGL model, the above subspaces have dimensions 4096, 4096, 4096, 4086, 4081, 4065, 2079, 2077, 2068, and 2055, respectively. As a result, when we plot the half-chain entanglement entropy for all energy eigenstates of the model, the single-arch structure observed previously for periodic boundary conditions splits into two arches 
(See Fig.~\ref{figLQGL_open}(a)). The eigenstates belonging to the first six nearly equal-sized Krylov subspaces form the upper arch, while those from the remaining subspaces constitute the lower arch.\\

For $H^{\text{Tot}}$, the corresponding subspaces have dimensions 4096, 4096, 4096, 4096, 4096, 4096, 2080, 2080, 2080, and 2079. In this case, the entanglement entropy again exhibits prominent ripples, with more pronounced dips compared to the periodic boundary case. We present these behaviors in Fig.~\ref{figLQGL_open}(b).\\

\begin{figure}
  \includegraphics[width=0.6\textwidth]{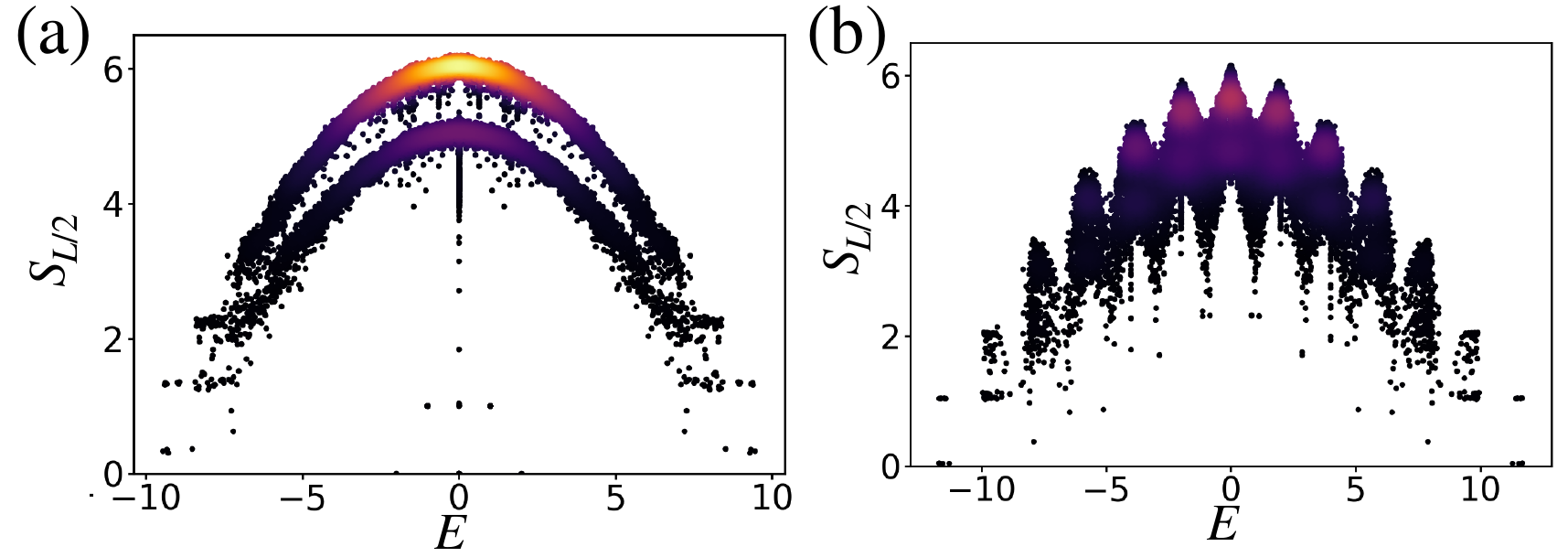}
  \caption{Behavior of half-chain entanglement entropy $S_{L/2}$ obtained for all eigenstates of (a) quantum game of life model. Here we have considered open boundary conditions and the eigenstates of the first six Krylov subspaces form the first arch (dimensions:  4096, 4096, 4096, 4086, 4081, and 4065). Whereas the second arch is formed by the eigenstates from the remaining four subspaces (dimensions: 2079, 2077, 2068, and 2055). Similar behavior is shown in  (b) for  $H^{\text{Tot}}$. Here the dimensions of the first biggest sized Krylov spaces are given by 4096, 4096, 4096, 4096, 4096, 4096, 2080, 2080, 2080,
and 2079. In both the models,  we consider $L=16$.}
  \label{figLQGL_open}
\end{figure}

\section{Choice of symmetry sector beyond $k=0$}
\label{sec_k_other}

\begin{figure}[htbp]
  \includegraphics[width=0.75\textwidth]{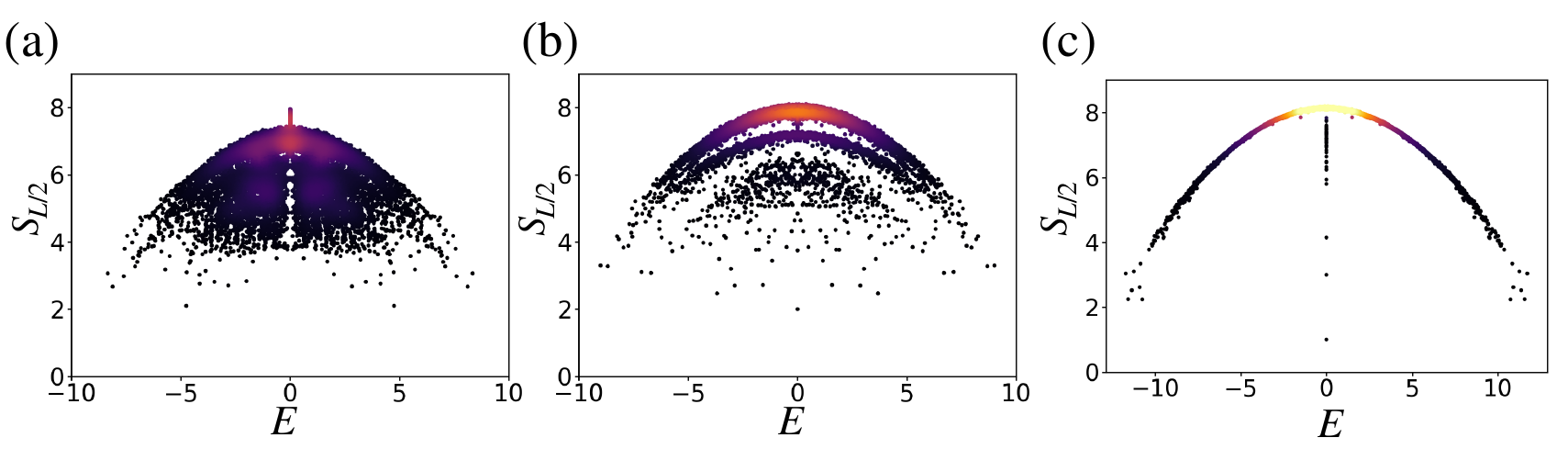}
  \caption{Behavior of half-chain entanglement entropy obtained for all eigenstates of (a) $H^{\mathcal{N}(1)}$, (b) $H^{\mathcal{N}(2)}$, and (c)  $H^{\text{QGL}}$. Here we have considered $k=1$ subspace with periodic boundary conditions and $L=18$. The behaviors show no qualitative difference from similar figures obtained for the $k=0$ subspaces.} 
 \label{figH1H2_k_1} 
\end{figure}

We now analyze whether the choice of momentum sectors beyond $ k=0$ shows similar behavior or not. Here we report that the effects of choosing Krylov subspaces other than $k=0$ (we have chosen $k=1$), is insignificant for the main results of our work, as long as the dimensions are comparable. While certain quantitative aspects—such as the size of the largest Krylov sectors do change, the qualitative features remain unchanged. As an illustration, we present the behavior obtained for the Hamiltonians $ H^{\text{QGL}}, H^{\mathcal{N}(1)}$ and $H^{\mathcal{N}(2)}$ in Fig.~\ref{figH1H2_k_1}. \\

\section{Krylov subspace structures of $H^{\mathcal{N}(0)}$}
\label{sec:fragH0}
\begin{figure}[htbp]
    \centering
    \includegraphics[width=0.6\textwidth]{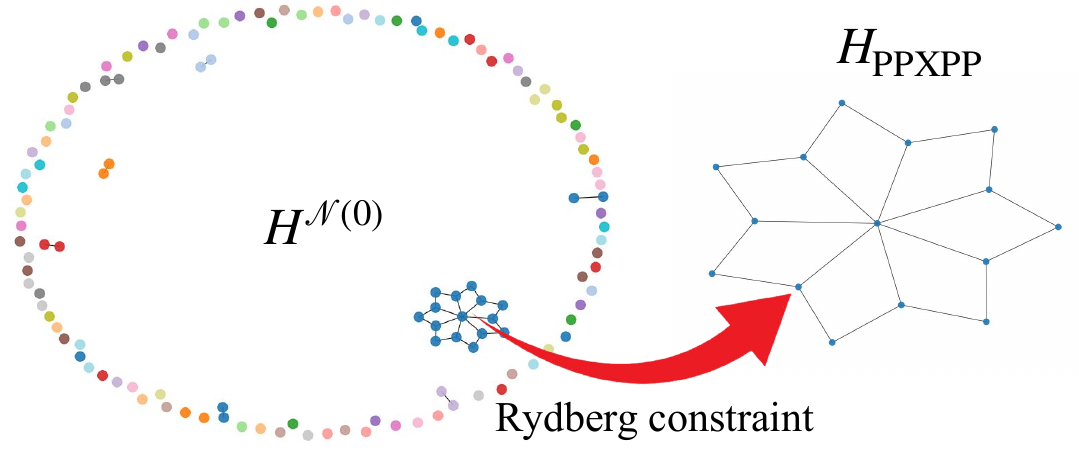} 
    \caption{In this schematic, we present the fragmentation of the Hilbert space for 
$H^{\mathcal{N}(0)}$ with $L=7$. Most of the Krylov subspaces consist of a 
single basis state that is annihilated by $H^{\mathcal{N}(0)}$. When we further 
impose the Rydberg constraint---namely, that basis states cannot contain any 
configuration other than $|\dots 00100 \dots\rangle$, more states get  eliminated, leading to the disappearance of all Krylov subspaces except for the 
large one containing 15 basis states and this action is 
indicated by the red arrow.}
    \label{fig_fragH0}
\end{figure}

In this section, in Fig.~\ref{fig_fragH0}, we schematically illustrate the Krylov subspaces of $H^{\mathcal{N}(0)}$ for a 
small system size $L=7$ and show that most of them are trivial, consisting of states that are annihilated by the Hamiltonian. We further show that the Rydberg constraint eliminates other all but the largest Krylov subspace, with 15 basis states which are the basis states in PPXPP model (with Rydberg constraint). These states are $|0000000\rangle, \ |1000000\rangle, \ |0100000\rangle, \ |0010000\rangle, \ |0001000\rangle, \ |0000100\rangle, \ |0000010\rangle, \ |0000001\rangle, \ |1001000\rangle, \ |0100100\rangle$, $ |0010010\rangle, \ |0001001\rangle, \ |1000100\rangle, \ |0100010\rangle \ \text{and} \ |0010001\rangle$. On the other hand, $|1100000\rangle$ and $|1100100\rangle$ form a Krylov subspace in the Hilbert space of $H^{\mathcal{N}(0)}$, but gets eliminated in PPXPP model due to Rydberg constraint. As a result, 
the conventional PPXPP model does not exhibit any fragmentation. Note that for large $L$, we can have a significant number of Krylov subspaces (nontrivial) with moderate dimension.


\section{Description of unfolding process of energy spectrum }
\label{sec:ulfolding}
\begin{figure}[htbp]
    \centering
    \includegraphics[width=0.6\textwidth]{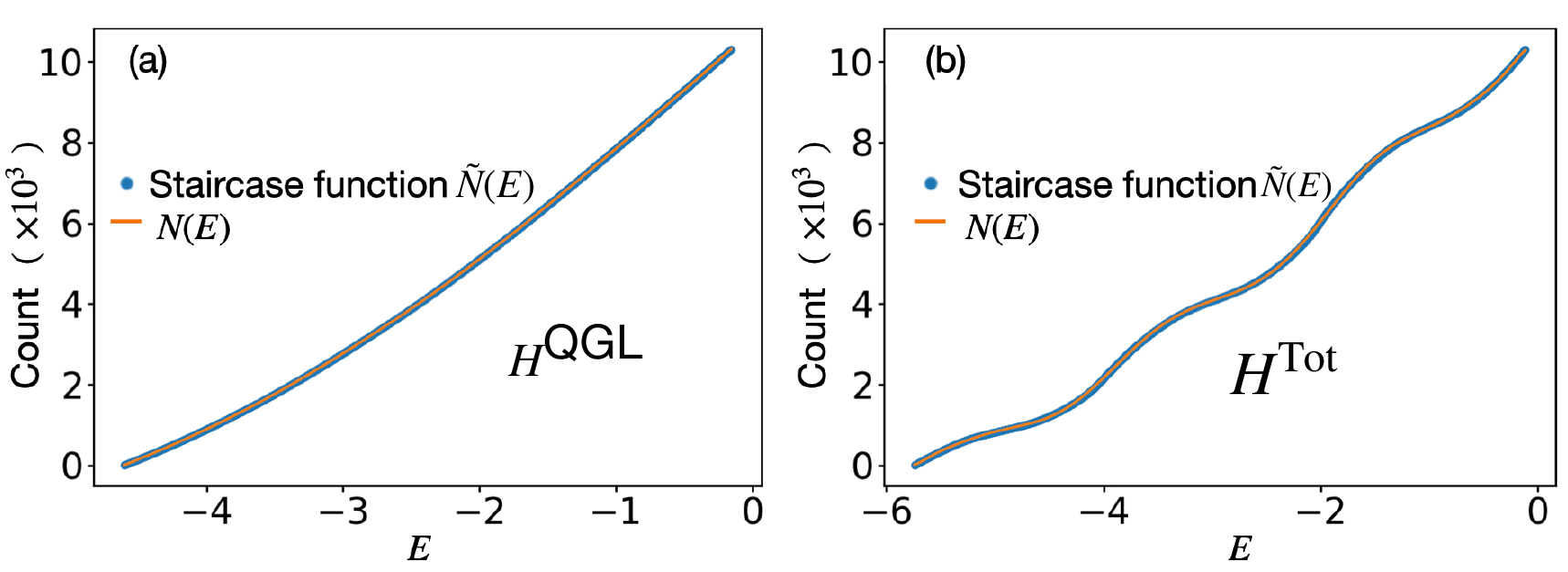} 
    \caption{(a) The staircase function $\widetilde{N}(E)$ and corresponding fitted polynomial ${N}(E)$ for $H_{\text{QGL}}$. (b) Shows similar plot   for $H^{\text{Tot}}$. We can see the imprint of the localization kind of property of $H^{\text{Tot}}$ is also visible here. Because of this particular plot, we had to use a high-degree polynomial $(15\text{th order} )$ for better fitting, which we used in all other Hamiltonians to maintain uniformity.}
\end{figure}

The unfolding process involves the following steps.  At first, we create a `staircase function' ($\widetilde{N}(E)$) that counts the number of energy levels that lie within $[E_{\mathrm{min}},E]$. To make this function, we do not use the full energy spectrum of a symmetry-reduced sector. If the total number of eigenvalues in a sector is \( D \), the energy list \( \{E_i\} \) is used, where \( i \) ranges from \( [D/10] \) to \( [D/2 - 500] \). So in $x$-axis, we plot the energy values of the mentioned energy range and in the $y$-axis, the corresponding cumulative count of energies till that energy value, $\widetilde{N}(E)$.  Thereafter, we fit that data with a polynomial function. Here, we have used a polynomial of order 15. Let's denote the fitted polynomial function as $N(E)$.  Then the unfolded spectrum is given by $\{e_n\} = \{N(E_n)\}$.

\section{Spin-flip symmetry resolutions for $H^{\mathcal{N}(2)}$}
\label{sec:h2_extra}

\begin{figure}[htbp]
    \centering
    \includegraphics[width=0.6\textwidth]{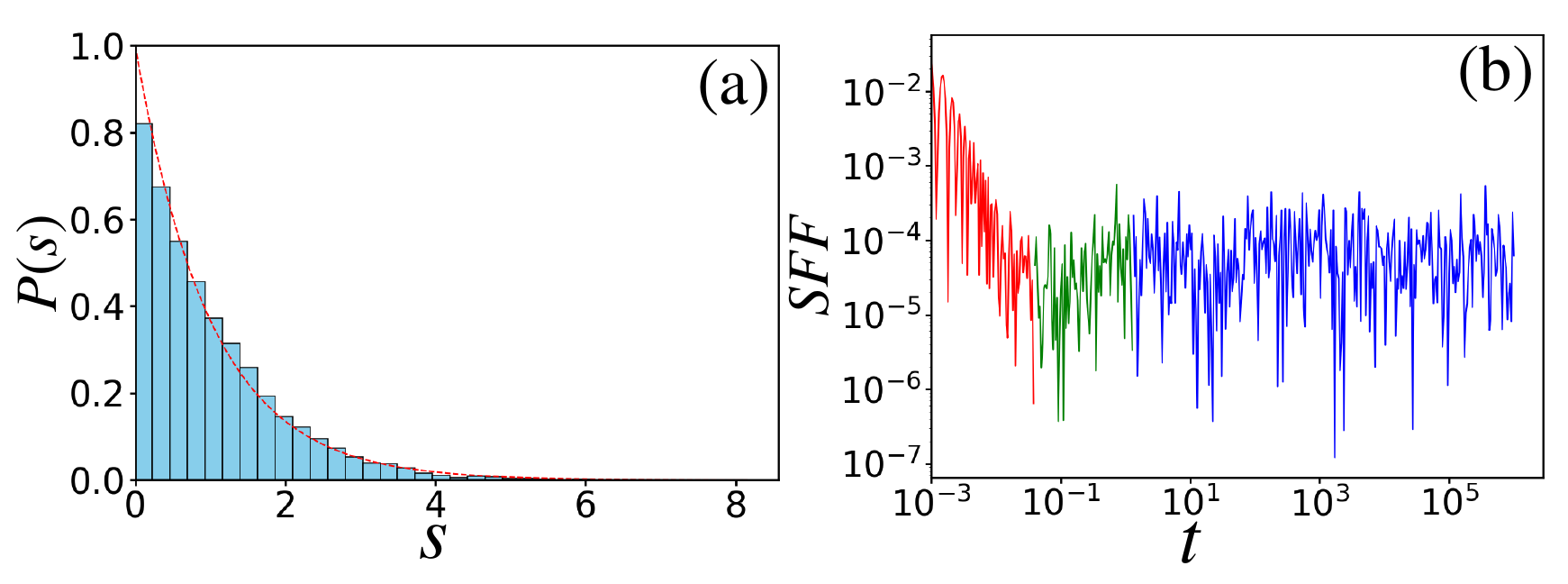} 
    \caption{(a) Level spacing distribution for $H^{\mathcal{N}(2)}$ in zero momentum inversion symmetric sector before considering fragmentation with L=20. (b) SFF for the same.}
    \label{fig:H_2_unresolved}
\end{figure}
In the Hamiltonian $H^{\mathcal{N}(2)}$, when we resolve all three (translation, inversion and spin-flip) symmetries, and take the biggest fragmented subspace, we get to see the Wigner-Dyson type level spacing distribution. However, if we just resolve the translation and inversion symmetry and consider all fragments together, surprisingly, a Poisson-type curve arises, which may give a notion of pseudo-integrability. However, for this case, when we plot the SFF, we get to see a clear ramp, which points towards the hidden chaotic nature. Only after resolving the remaining symmetry and taking into account the fragmentation, we get the Wigner-Dyson distribution here. This strongly suggests that in symmetry resolution, SFF provides better diagnostics than level spacing distribution.

\end{widetext}

\bibliography{reference}{}

\end{document}